\documentclass[journal,twoside]{IEEEtran}
\usepackage[english]{babel}
\usepackage[latin1]{inputenc}
\usepackage{graphicx}
\usepackage{subfigure}
\usepackage{amssymb}
\usepackage{amsmath}
\usepackage{verbatim}
\usepackage{cite}
\usepackage{newalg}
\usepackage{url}

\begin{document}

\newtheorem{thm}{Theorem}
\newtheorem{cor}{Corollary}
\newtheorem{lem}{Lemma}
\newtheorem{dfn}{Definition}
\newcommand{\argmin}{\mbox{arg\,min}}

\title{Polynomial-Time Algorithms for Multirate Anypath Routing in Wireless Multihop Networks}

\author{Rafael Laufer, Henri Dubois-Ferri\`ere, and Leonard Kleinrock%
\thanks{This work was supported by the U.S. National Science Foundation under Grants NBD-0721963 and CCF-0120778, and by a gift from Bell Labs, Alcatel-Lucent Foundation. A preliminary version of this paper was published in the Proceedings of IEEE INFOCOM 2009.
}
\thanks{Rafael Laufer and Leonard Kleinrock are with the Computer Science Department, University of California, Los Angeles, Los Angeles, CA 90095 USA (email: rlaufer@cs.ucla.edu). Henri Dubois-Ferrière is with Riverbed Technology, Inc., Lausanne, Switzerland.}%
\\
September 17, 2010\\
Technical Report UCLA-CSD-TR100034
\vspace{-.5in}
}

\maketitle

\begin{abstract}
In this paper we present a new routing paradigm that generalizes opportunistic routing for wireless multihop networks. In multirate anypath routing, each node uses both a set of next hops and a selected transmission rate to reach a destination. Using this rate, a packet is broadcast to the nodes in the set and one of them forwards the packet on to the destination. To date, there has been no theory capable of jointly optimizing both the set of next hops and the transmission rate used by each node. We solve this by introducing two polynomial-time routing algorithms and provide the proof of their optimality. The proposed algorithms run in roughly the same running time as regular shortest-path algorithms, and are therefore suitable for deployment in routing protocols. We conducted measurements in an 802.11b testbed network, and our trace-driven analysis shows that multirate anypath routing performs on average 80\% and up to 6.4 times better than anypath routing with a fixed rate of 11~Mbps. If the rate is fixed at 1~Mbps instead, performance improves by up to one order of magnitude.
\end{abstract}


\section{Introduction}

\IEEEPARstart{R}{outing} in wireless multihop networks is challenging due to the high loss rate and dynamic quality of wireless links~\cite{campista08,fernandes09,velloso10,passos09}. Anypath routing\footnote{We prefer the term {\it anypath routing} instead of {\it opportunistic routing}, since opportunistic routing is an overloaded term also used for opportunistic contacts~\cite{seth07}. 
We explain the meaning of the word anypath in Section~\ref{sec:anypath}.} has been recently proposed as a way to circumvent these shortcomings by using multiple next hops for each destination~\cite{chachulski07b,biswas05a,zhong06a,dubois-ferriere10}. Each packet is broadcast to a forwarding set composed of several neighbors, and the packet is lost only if none of these neighbors receive it. Therefore, while the link to a given neighbor is down or performing poorly, another nearby neighbor may receive the packet and forward it on. This is in contrast to single-path routing where only one neighbor is assigned as the next hop for each destination. In this case, if the link to this neighbor is poor, a packet may be lost even though other neighbors may have overheard it.

Existing work on anypath routing has focused on wireless networks that use a single transmission rate. However, certain wireless systems offer multiple transmission rates at the physical layer; this is notably the case for 802.11a/b/g/n. For these physical layers, restricting multihop communication to a single bit rate means that routing decisions cannot take advantage of different modulation/coding schemes and the associated tradeoff. This presents at least two drawbacks.
First, using a single rate over the entire network underutilizes available bandwidth resources. Some links may perform well at a higher rate, while others may only work at a lower rate. Second, and most importantly, the network may become disconnected at a higher bit rate. We provide experimental measurements from an 802.11b testbed which show that this phenomenon is not uncommon in practice. The key problem is that higher bit rates have a shorter transmission range, which reduces the network connectivity. As the rate increases, links becomes lossier and the network eventually gets disconnected. Therefore, in order to guarantee connectivity, single-rate anypath routing must be limited to low rates.

In {\it multirate anypath routing}, these problems do not exist; however, we face different challenges. First, loss probabilities increase with higher transmission rates, so a higher bit rate does not always improve throughput. Second, we must find not only the forwarding set, but also the transmission rate at each hop that jointly minimizes its cost to a destination. For instance, assuming that links $(i,j)$, $(i,k)$, and $(i,l)$ achieve their highest throughput at 2, 5.5, and 11~Mbps, respectively, which subset of neighbors should node~$i$ use to reach the destination and at which rate should the packet be transmitted? Finally, higher rates have a shorter transmission range and therefore we have a different connectivity graph for each rate. Lower rates have more neighbors available for inclusion in the forwarding set (i.e., more spatial diversity) and fewer hops between nodes. Higher rates have fewer neighbors available for the forwarding set (i.e., less spatial diversity) and longer routes~\cite{kleinrock75}. Finding the optimal operation point in this tradeoff is the focus of this paper. 

We thus address the problem of finding both a forwarding set and a transmission rate for every node, such that the overall cost of every node to a particular destination is minimized. We call this the {\it shortest multirate anypath problem}. To our knowledge, this is still an open problem~\cite{biswas05a,chachulski07b,zeng08}, and we believe our algorithms are the first practical solution for it.

e introduce two polynomial-time routing algorithms to the shortest multirate anypath problem and present a proof of their optimality. Our solution generalizes Dijkstra's and Bellman-Ford algorithms for the multirate anypath case and are applicable to both link-state and distance-vector routing protocols, respectively. One would expect the running time of such algorithms to be exponential, since with $n$ neighbors we can have up to $2^n-1$ forwarding sets. However, we show that the proposed algorithms have roughly the {\it same} polynomial time as the corresponding shortest-path algorithms, and are suitable for implementation at current wireless routers. We also provide a generalization of the expected transmission time (ETT) routing metric~\cite{draves04b} for multirate anypath routing. 

To evaluate performance, we conducted measurements in an 18-node 802.11b wireless testbed of embedded Linux devices. Our results reveal that the network becomes partially disconnected if we fix the transmission rate of every node at 2, 5.5, or 11~Mpbs. A single-rate routing scheme therefore performs poorly in this case, since 1~Mbps is the only rate at which the network is fully connected. Using a trace-driven analysis, we show that multirate anypath routing improves the end-to-end expected transmission time by 80\% on average and up to 6.4 times compared to single-rate anypath routing at 11~Mbps, while still maintaining full network connectivity. The performance is even higher for the single-rate case at 1~Mbps, with an average gain of a factor of 5.4 and a maximum gain of a factor of 11.3.

The remainder of this paper is organized as follows. Section~\ref{sec:anypath} reviews the basic theory of anypath routing and our network model and assumptions. In Section~\ref{sec:multirate}, we introduce multirate anypath routing and the proposed routing metric. Section~\ref{sec:shortest} presents the multirate anypath algorithms and proves their optimality. Section~\ref{sec:results} reveals the results of our evaluation, showing the benefits of multirate over single-rate anypath routing. Section~\ref{sec:related} presents the related work in anypath routing. Finally, conclusions are presented in Section~\ref{sec:conclusions}.

\vspace{-0.25cm}
\section{Anypath Routing}
\label{sec:anypath}

In this section we review the anypath routing theory introduced by Zhong~{\it et~al.}~\cite{zhong06a} and Dubois-Ferrière~{\it et~al.}~\cite{dubois-ferriere10}. We also generalize this theory to support correlated link losses.

\vspace{-0.15in}
\subsection{Overview}

In classic wireless network routing, each node forwards a packet to a single next hop. As a result, if the transmission to that next hop fails, the node needs to retransmit the packet even though other neighbors may have overheard it. In contrast, in anypath routing, each node broadcasts a packet to {\it multiple} next hops simultaneously. Therefore, if the transmission to one neighbor fails, an alternative neighbor who received the packet can forward it on. We define this set of multiple next hops as the {\it forwarding set}, and we usually use~$J$ to represent it throughout the paper. A different forwarding set is used to reach each destination, in the same way that a distinct next hop is used for each destination in classic routing. 

When a packet is broadcast to the forwarding set, more than one node may receive the same packet. To avoid unnecessary duplicate forwarding, only one of these nodes should forward the packet on. For this purpose, each node in the set has a priority in relaying the received packet. A node only forwards a packet if all higher priority nodes in the set failed to do so. Higher priorities are assigned to nodes with lower costs to the destination. As a result, if the node with the lowest cost in the forwarding set successfully received the packet, it forwards the packet to the destination while others suppress their transmission. Otherwise, the node with the second lowest cost forwards the packet, and so~on. A reliable anycast scheme~\cite{jain08} is necessary to enforce this relay priority. We talk more about this in Section~\ref{sec:model_assumptions}. The source keeps rebroadcasting the packet until someone in the forwarding set receives it or a threshold is reached. Once a neighbor in the set receives the packet, this neighbor repeats the same procedure until the packet is delivered to the destination.

\begin{figure}[ht!]
\centering
\includegraphics[width=.35\textwidth]{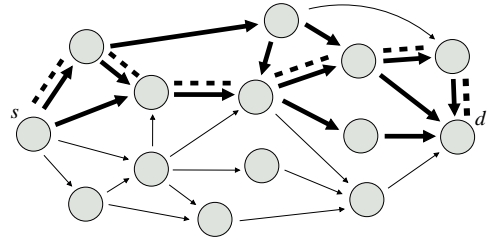}
\caption{An anypath connecting nodes $s$ and $d$ is shown in bold arrows. The anypath is composed of the set of 11 paths between the two nodes. Every packet sent from $s$ traverses one of these paths to reach $d$, such as the path shown with a dashed line. Different packets may traverse different paths, depending on which nodes receive the forwarded packet at each hop; hence the name anypath.}
\label{fig:anypath}
\vspace{-0.1in}
\end{figure}

Since we now use a set of next hops to forward packets, every two nodes are connected through a mesh composed of multiple paths. Fig.~\ref{fig:anypath} depicts this scenario where each node uses a set of neighbors to forward packets. The forwarding sets are defined by the multiple bold arrows leaving each node. We define this set of paths between two nodes as an {\it anypath}. In the figure, the anypath shown in bold is composed by 11 different paths between a source~$s$ and a destination~$d$. Depending on the choice of each forwarding set, different paths are included in or excluded from the anypath. At every hop, only a single node of the set forwards the packet on. Consequently, every packet from~$s$ traverses only one of the available paths to reach~$d$. We show a path possibly taken by a packet using a dashed line. Succeeding packets, however, may take completely different paths; hence the name anypath. The path taken is determined on the fly, depending on which nodes of the forwarding set successfully receive the packet at each hop.

\vspace{-0.1in}
\subsection{System Model and Assumptions}
\label{sec:model_assumptions}

In order to support the point-to-multipoint links used in anypath routing, we model the wireless mesh network as a hypergraph. A hypergraph $\mathcal{G}=(V,\mathcal{E})$ is composed of a set~$V$  of vertices or nodes and a set~$\mathcal{E}$ of hyperedges or hyperlinks. A hyperlink is an ordered pair $(i,J)$, where $i \in V$ is a node and~$J$ is a nonempty subset of~$V$ composed of neighbors of~$i$. For each hyperlink $(i,J) \in \mathcal{E}$, we have a delivery ratio~$p_{iJ}$ and a cost~$d_{iJ}$. If the set $J$ has a single element~$j$, then we just use $j$ instead of $J$ in our notation. In this case, $p_{ij}$ and~$d_{ij}$ denote the link delivery ratio and the link cost, respectively. 

The hyperlink delivery ratio $p_{iJ}$ is defined as the probability that a packet transmitted from~$i$ is successfully received by at least one of the nodes in~$J$. If $J = \{1,2,\ldots,n\}$, then $p_{iJ}$ is
\begin{equation}
\label{eq:p_iJ}
p_{iJ} = 1-P[\,X_1 = 0, X_2 = 0, \ldots, X_n = 0\,],
\end{equation}
where $X_j$ is a random variable equal to 0 or 1 if node~$j \in J$ loses or receives a packet transmitted by~$i$, respectively.
If packet losses occur independently at different receivers, as in light load regimes (Section~\ref{sec:results}), then the ratio $p_{iJ}$ is simplified to 
$1-\prod_{j \in J} \left(1-p_{ij}\right)$,
which can be calculated from the individual link delivery ratios~$p_{ij}$. 
For high network loads, however, interference may induce some correlation among neighbor links, and 
this may not be a perfect estimate of~$p_{iJ}$. 
Nonetheless, regardless of correlation or independence, the ratio $p_{iJ}$ can always be locally estimated at node~$i$ using active (e.g., periodic probing and reporting to neighbors) or passive (e.g., unackownledged data frames) measurements, which should be performed periodically for better accuracy. Since $p_{iJ}$ is basically the fraction of successful MAC frames transmitted to~$J$, it already takes into account the effect of path loss and interference on packet reception. Similar approaches for estimating delivery ratios are used in~\cite{couto03,draves04b,reis06,biswas05a}.

We experimentally estimate the delivery ratios of our testbed in Section~\ref{sec:results}. Nonetheless, for ease of presentation, we assume in our examples that link loss probabilities are known and independent at each receiver. The independence assumption, however, is not necessary for our routing algorithms to work; our optimality proofs are general and also hold for the case of correlated losses. This is in contrast to previous work on anypath routing~\cite{chachulski07b,biswas05a,zhong06a,dubois-ferriere10}, which always assumes link losses are independent, even though some correlation may indeed occur in practice~\cite{srinivasan10,zhu10}.

Previously proposed MAC protocols have been designed to guarantee the relay priority among the nodes in the forwarding set~\cite{biswas05a,zorzi03,jain08}. Such protocols can use different strategies for this purpose, such as time-slotted access, prioritized contention, and frame overhearing. Reliable anycast is an active area of research, and we assume that such a mechanism is in place to ensure that the relaying priority is respected. The details of the MAC, however, are abstracted from the routing layer. Practical routing protocols only incorporate the delivery ratios into the routing metric in order to abstract from the MAC details~\cite{couto03,draves04b} and we take the same approach. The only MAC aspect that is important is the effectiveness of the relaying node selection.

\vspace{-0.1in}
\subsection{Anypath Cost}
\label{sec:anypath_cost}

In this section we explain how to calculate the anypath cost from a node~$i$ to a given destination via a forwarding set~$J$. The anypath cost $D_i$ is defined as $D_i = d_{iJ} + D_J$, which is composed of the hyperlink cost $d_{iJ}$ from~$i$ to~$J$ and the remaining anypath cost $D_J$ from~$J$ to the destination. We now explain each one of these costs in detail.

The hyperlink cost $d_{iJ}$ depends on the routing metric used. 
Most of the previous works on anypath routing adopted the expected number of anypath transmissions (EATX) as the routing metric~\cite{zhong06a, chachulski07b, dubois-ferriere10}. The EATX is a generalization of the unidirectional ETX metric~\cite{couto03}, which is defined as $d_{ij} = 1/p_{ij}$. The cost~$d_{ij}$ for ETX represents the expected number of transmissions necessary for a packet sent by~$i$ to be successfully received by~$j$. This metric generalizes the previous hop count metric by including the link quality into the cost. For EATX, the cost~$d_{iJ}$ is defined as $d_{iJ} = 1/p_{iJ}$, which is the average number of transmissions necessary for at least one node in~$J$ to correctly receive the transmitted packet.

The remaining anypath cost $D_J$ is defined as a {\it weighted average} of the costs of the nodes in the forwarding set 
\begin{equation}
\label{eq:D_J_def} D_J = \sum_{j \in J} w_{ij}D_j, \mathrm{\ with\ }\sum_{j \in J} w_{ij} = 1,
\end{equation}
where the weight~$w_{ij}$ in~(\ref{eq:D_J_def}) is the probability of node~$j$ being the relaying node of a packet from node~$i$. For example, let the forwarding set be defined as $J = \{1,2,\ldots,n\}$ with costs $D_1 \leq D_2 \leq \ldots \leq D_n$. 
Node~$j$ will be the relaying node only when it receives the packet and none of the nodes with a lower cost also receive it, which happens with probability 
$P[\,X_1=0,\ldots,X_{j-1}=0,X_j=1\,]$.
The weight~$w_{ij}$ is then 
\begin{equation}
\label{eq:w_j_def} w_{ij} = \frac{P[\,X_1=0,\ldots,X_{j-1}=0,X_j=1\,]}{1-P[\,X_1=0,X_2=0,\ldots,X_n=0\,]}, 
\end{equation}
with the denominator being the normalizing constant. The weight~$w_{ij}$ in (\ref{eq:w_j_def}) can also be interpreted as the probability of node~$j$ being the relaying node, given that at least one of the nodes in~$J$ received the packet. For the case of independent losses, the weight~$w_{ij}$ can be simplified to
\vspace{-0.05in}
\begin{equation}
w_{ij} = \frac{\displaystyle p_{ij}\prod_{k=1}^{j-1}\left(1-p_{ik}\right)}{1-\displaystyle\prod_{j \in J} \left(1-p_{ij}\right)}.
\end{equation}
\vspace{-0.05in}

\begin{figure}[ht!]
\centering
\subfigure[]{
  \label{fig:example-a}
  \includegraphics[width=.20\textwidth]{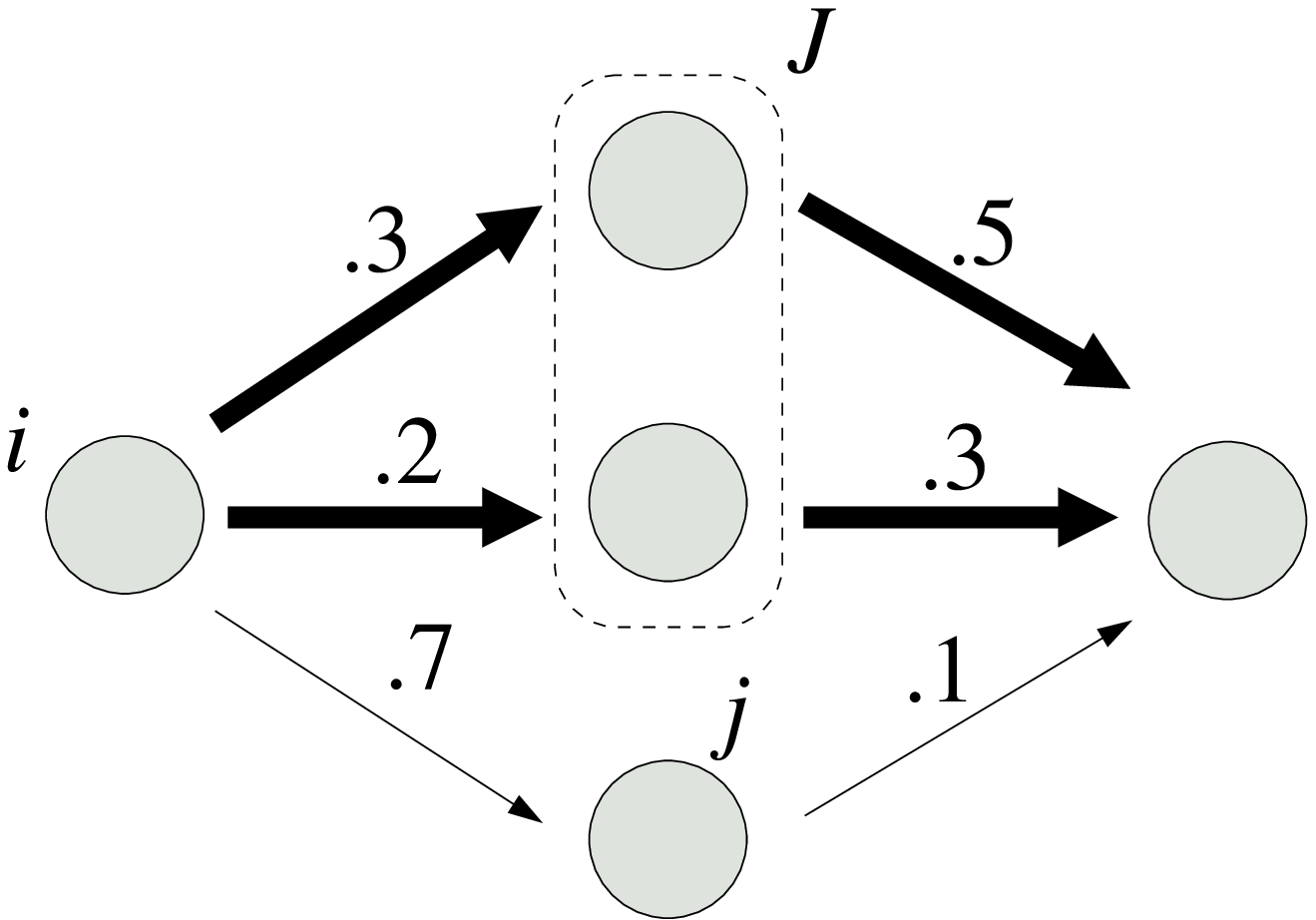}
}
\subfigure[]{
  \label{fig:example-b}
  \includegraphics[width=.20\textwidth]{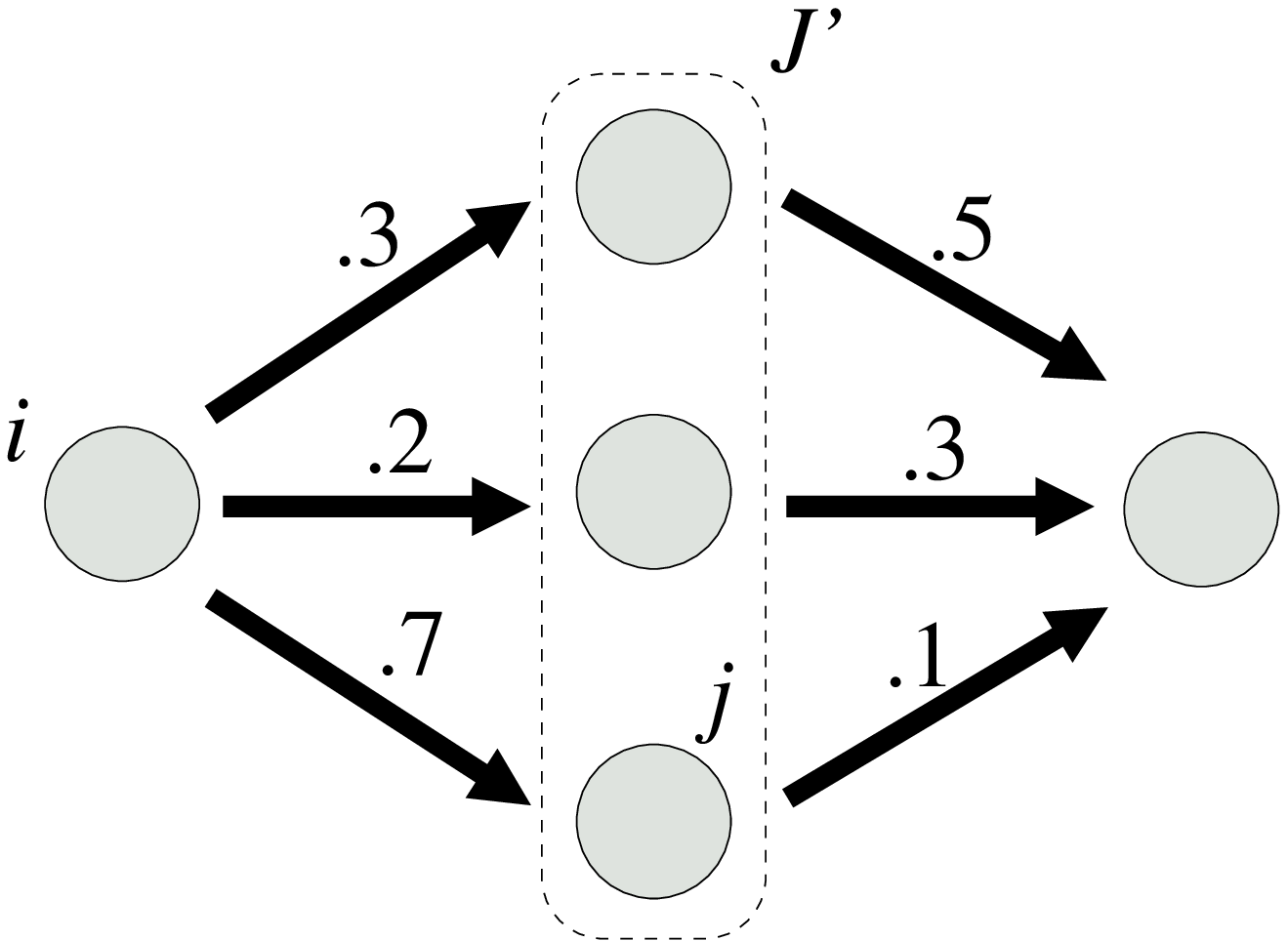}
}
\vspace{-0.1in}
\caption{An anypath cost calculation example. 
The weights represent the link delivery ratios. The anypath cost in~(a) is lower than the cost in~(b).}
\label{fig:example}
\vspace{-0.1in}
\end{figure}

As an example, consider the network depicted in Fig.~\ref{fig:example}. Let~$J$ be the two-node forwarding set in Fig.~\ref{fig:example-a} and let~$J'$ be the three-node forwarding set in Fig.~\ref{fig:example-b}. 
The weight of each link represents the link delivery ratio, which is the inverse of the expected number of transmissions (ETX). 
The cost via~$J$ in Fig.~\ref{fig:example-a} is calculated as 
\setlength{\arraycolsep}{0.0em}
\begin{eqnarray}
\nonumber D_i &{}={}& d_{iJ} + D_J \\
\nonumber    &{}={}& \frac{1}{1-(1-0.3)(1-0.2)} + \frac{(0.3)2.0 + (1-0.3)(0.2)3.3}{1-(1-0.3)(1-0.2)} \\
             &{}={}& 2.3 + 2.4 = 4.7. 
\end{eqnarray}
\setlength{\arraycolsep}{5pt}%
One would expect that adding an extra node to the forwarding set is always beneficial because it increases the number of possible paths a packet can take. However, this is not always true, as shown in Fig.~\ref{fig:example-b}. The anypath cost via $J' = J \cup \{j\}$ is $D_i = d_{iJ'} + D_{J'} = 1.2 + 6.0 = 7.2$. On one hand, using~$J'$ instead of~$J$ reduces the hyperlink cost; that is, $d_{iJ'} \leq d_{iJ}$. On the other hand, the extra node increases the remaining anypath cost; that is, $D_{J'} \geq D_J$. If the increase $D_{J'}-D_J$ is higher than the decrease $d_{iJ} - d_{iJ'}$, adding this extra node is not worthwhile since the total cost to reach the destination increases. The intuition here is that when node~$j$ is the only one in~$J'$ to receive the packet, it is cheaper to retransmit the packet to one of the two nodes in~$J$ and take a shorter path from there than to take the long path via node~$j$.

\vspace{-0.1in}

Once the cost of an anypath is defined, we can find the anypath with the lowest cost; that is, the shortest anypath. This is called the {\it shortest-anypath problem}~\cite{dubois-ferriere10}. Interestingly, the shortest anypath will always have an equal or lower cost than the shortest single path. This is a direct consequence of the definition of an anypath as a set of paths. Among all possible anypaths between two nodes, we also have the anypath composed only of the path with the lowest cost. Therefore, if we are to choose the shortest anypath among all these possibilities, we know for sure that its cost can never be higher than the cost of the shortest single path.

\vspace{-0.05in}
\section{Multirate Anypath Routing}
\label{sec:multirate}

\begin{figure*}[ht!]
\centering
\subfigure[]{
  \label{fig:saf-execution-a}
  \includegraphics[width=.23\textwidth]{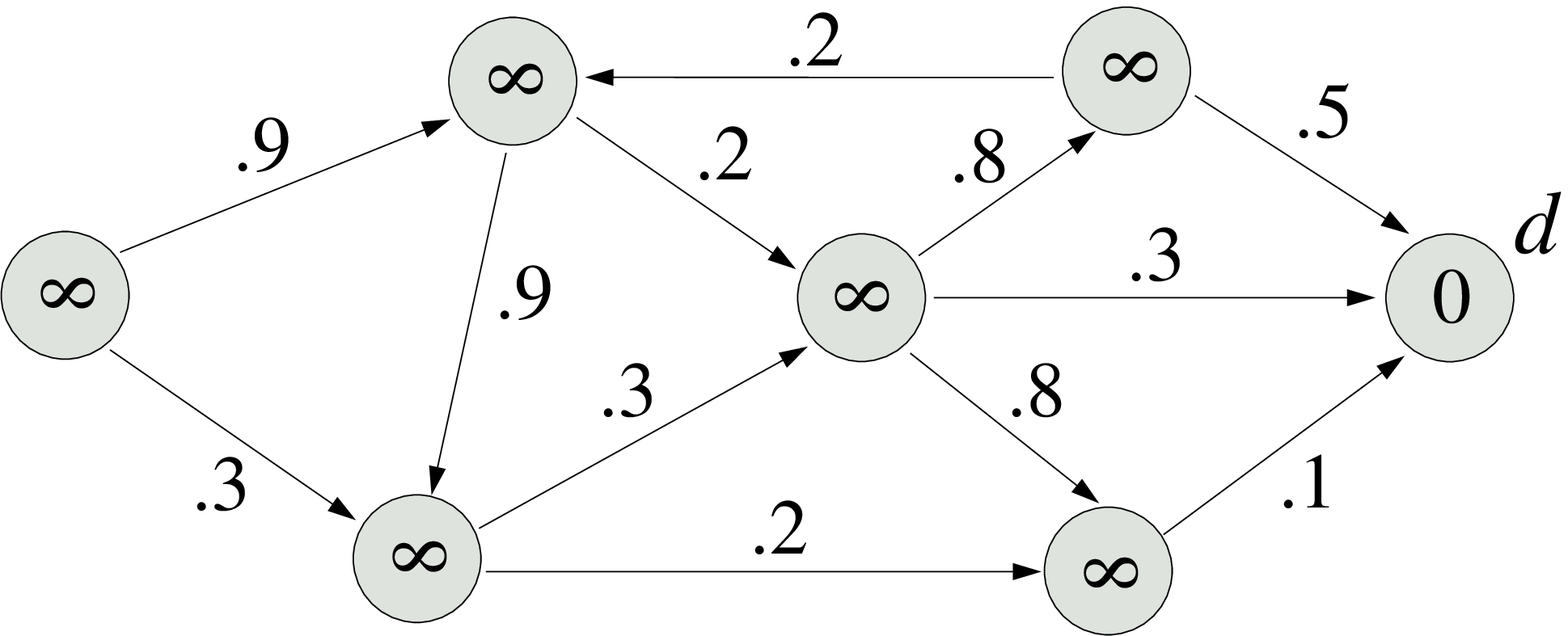}
}
\subfigure[]{
  \label{fig:saf-execution-b}
  \includegraphics[width=.23\textwidth]{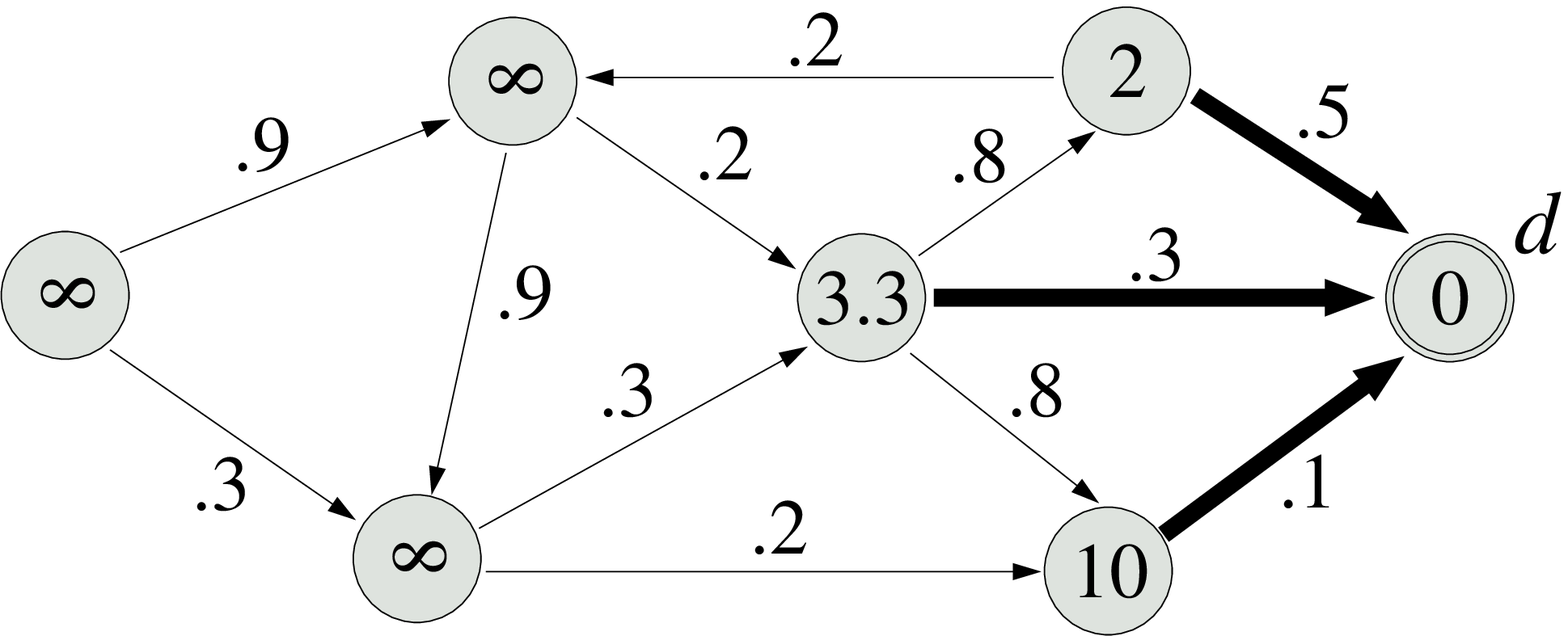}
}
\subfigure[]{
  \label{fig:saf-execution-c}
  \includegraphics[width=.23\textwidth]{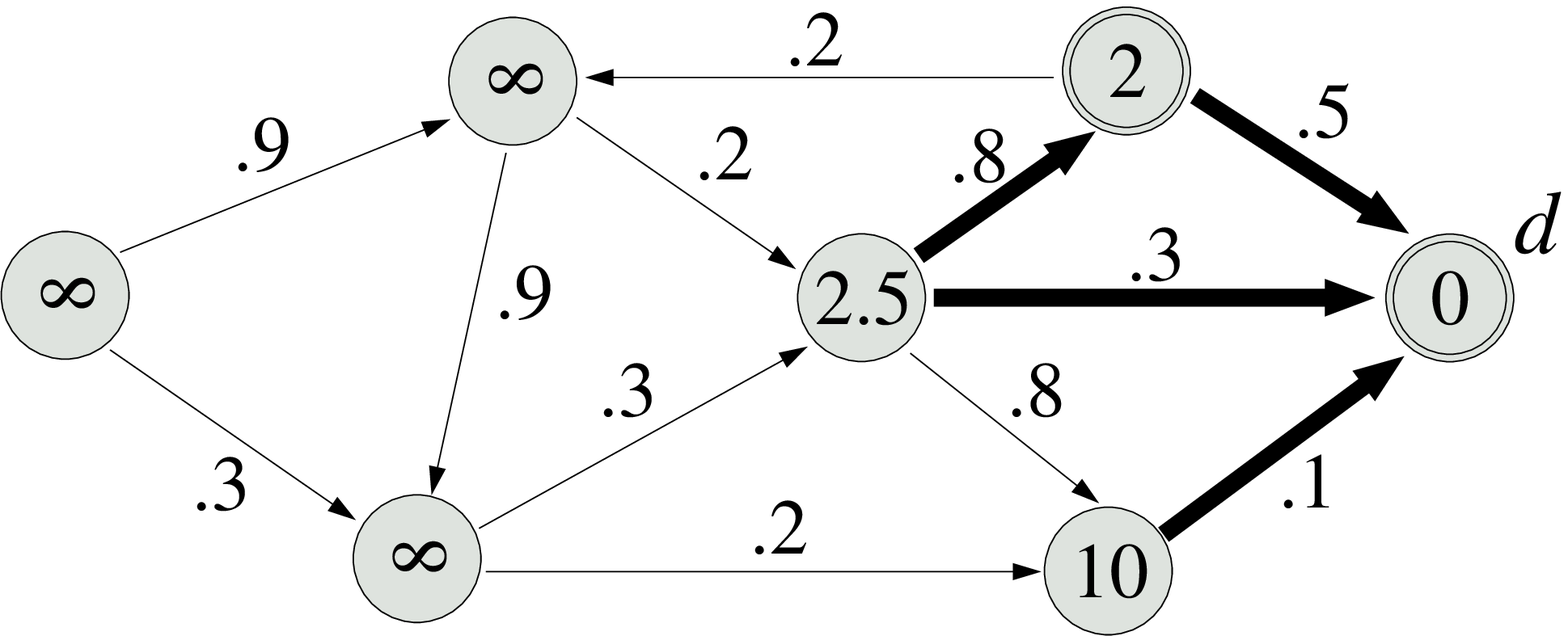}
}
\subfigure[]{
  \label{fig:saf-execution-d}
  \includegraphics[width=.23\textwidth]{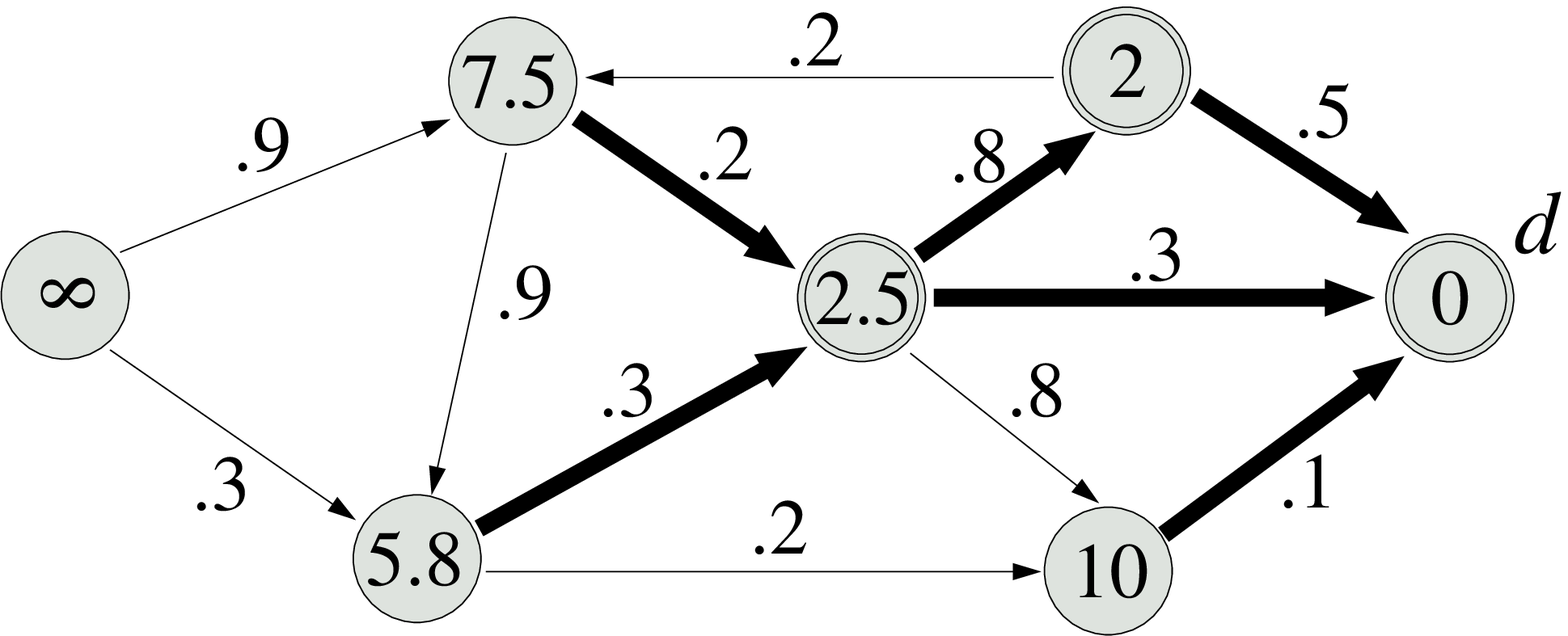}
}
\subfigure[]{
  \label{fig:saf-execution-e}
  \includegraphics[width=.23\textwidth]{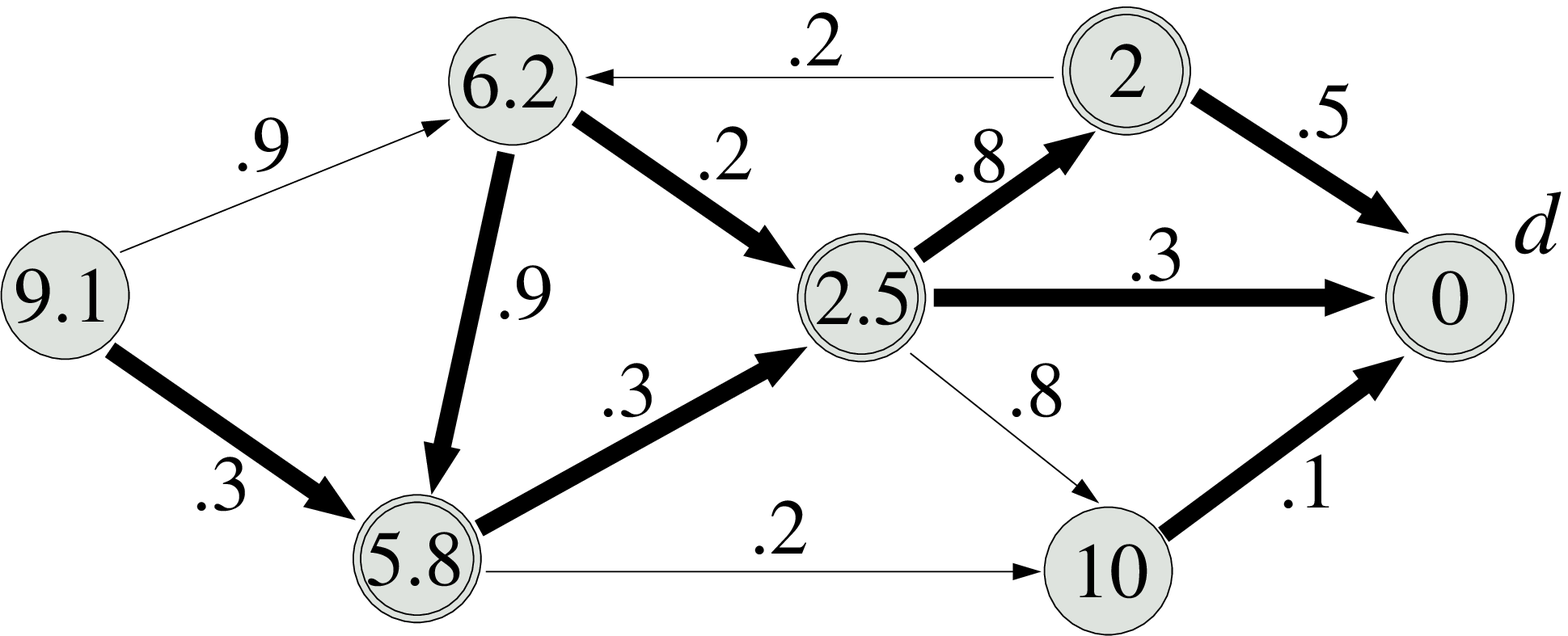}
}
\subfigure[]{
  \label{fig:saf-execution-f}
  \includegraphics[width=.23\textwidth]{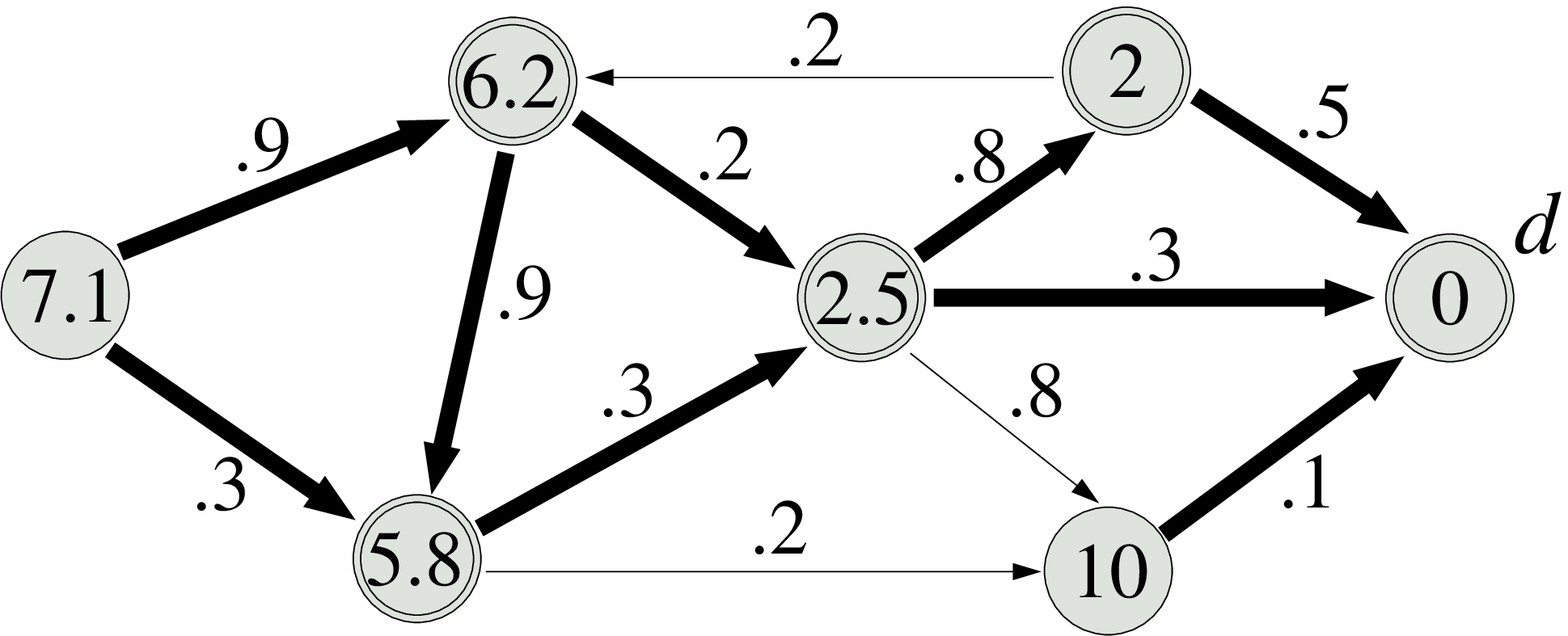}
}
\subfigure[]{
  \label{fig:saf-execution-g}
  \includegraphics[width=.23\textwidth]{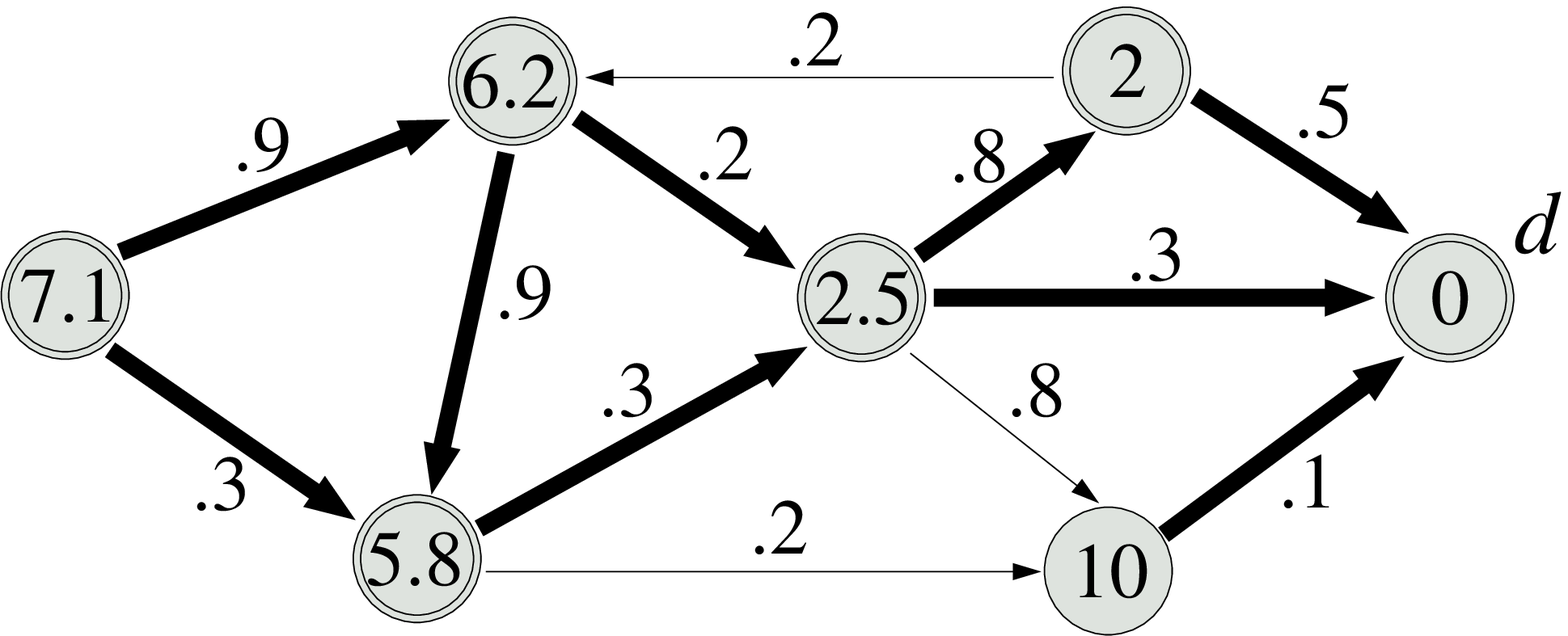}
}
\subfigure[]{
  \label{fig:saf-execution-h}
  \includegraphics[width=.23\textwidth]{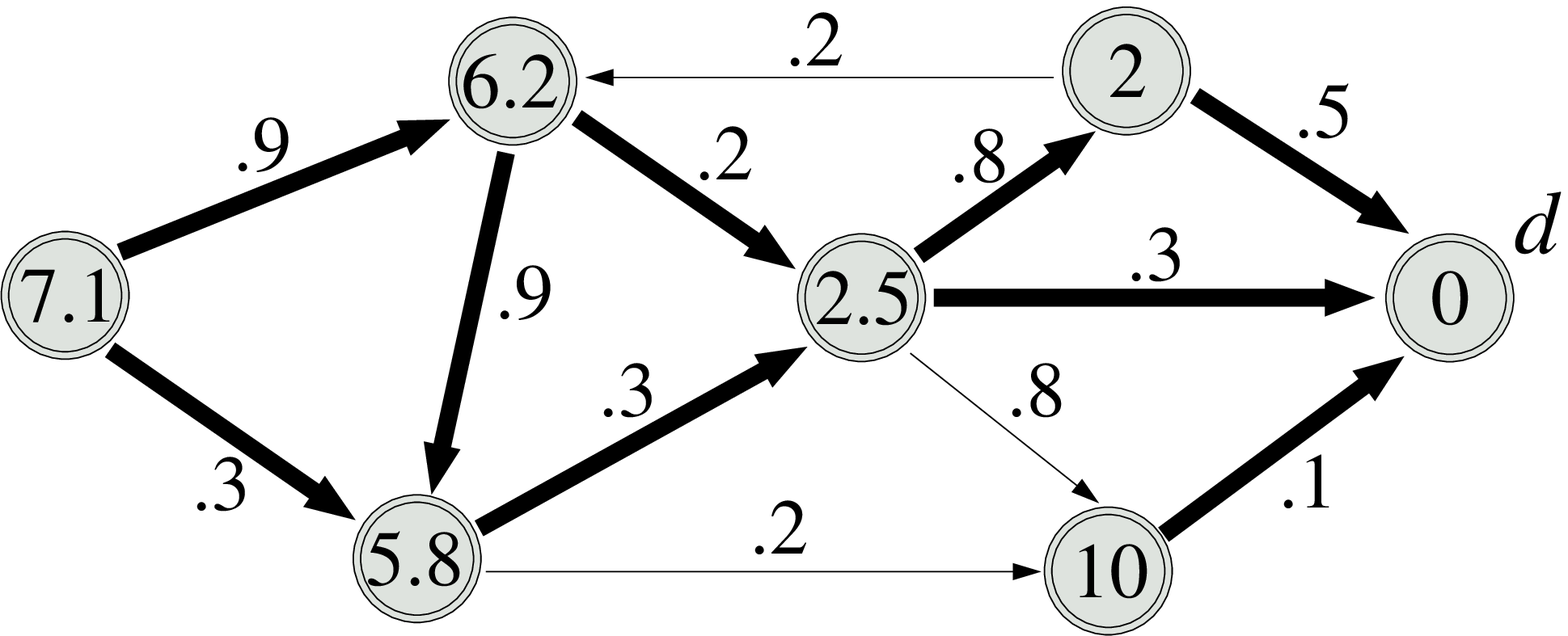}
}

\caption{Execution of the Shortest Anypath First (SAF) algorithm from every node to~$d$. The link weights represent the link delivery ratios, the value inside a node~$i$ is its cost $D_i$ to~$d$ (i.e., the end-to-end expected number of transmissions), and the arrows in boldface represent the shortest anypath to~$d$. (a) The situation just after the initialization. (b)--(g) The situation after each successive iteration of the algorithm. (h) The situation after the last node is settled.}
\label{fig:saf-execution}
\vspace{-0.1in}
\end{figure*}

\begin{figure}[ht!]
\centering
\includegraphics[width=.35\textwidth]{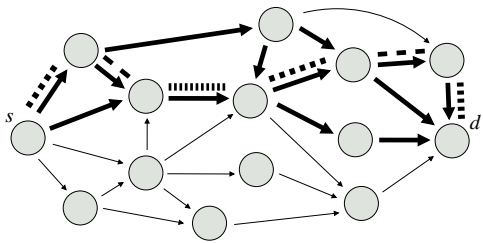}
\vspace{-0.05in}
\caption{A multirate anypath connecting nodes $s$ and $d$ is shown in bold arrows. Every packet sent from~$s$ traverses a path to reach~$d$, such as the path shown with dashed lines. Different dash lengths represent the different bit rates used by each node, with a shorter dash for higher rates. Altough not illustrated, the node degree and network topology change for different rates.}
\label{fig:multirate_anypath}
\vspace{-0.05in}
\end{figure}

Previous work on anypath routing focused on a single bit rate~\cite{biswas05a,zhong06a,chachulski07b,dubois-ferriere10}. Such an assumption, however, considerably underutilizes available bandwidth resources. Some hyperlinks may be able to sustain a higher transmission rate, while others may only work at a lower rate. Additionally, the transmission range and therefore the network topology change with the rate, which makes multirate a challenge for routing. To date, the problem of how to select the transmission rate for anypath routing is still open~\cite{chachulski07b,zeng08}. We provide a solution to this problem and incorporate the multirate capability inherent in 802.11 networks into anypath routing. In this case, besides selecting a set of next hops for packet forwarding, a node must also select a transmission rate. For each destination, a node then keeps both a forwarding set and a rate used to reach this set. As a result, every two nodes will be connected through a mesh composed of multiple paths, with each node transmitting at a selected rate. 
Fig.~\ref{fig:multirate_anypath} depicts this scenario.
We define this set of paths between two nodes, with each node using a potentially different bit rate, as a {\it multirate anypath}. In the figure, 
a packet is sent from~$s$ to~$d$ over the multirate anypath. Only one of the available paths is traversed, depending on which nodes successfully receive the packet at each hop. We show a path possibly taken by the packet using dashed lines. We use different dash lengths to represent the different 
rates used by each node. A shorter dash represents a shorter time to send a packet, hence a higher 
rate. Succeeding packets may take completely different paths with other 
rates along the way.

\vspace{-0.05in}

In order to support multirate, we must extend the system model in Section~\ref{sec:model_assumptions}. Let~$R$ be the set of available bit rates that nodes can use to transmit their packets. For each hyperlink $(i,J) \in \mathcal{E}$, we now have a delivery ratio~$p_{iJ}^{(r)}$ and a cost~$d_{iJ}^{\,(r)}$ associated with each transmission rate $r \in R$. In real wireless networks, we have different delivery ratios and costs for each transmission rate, which justifies this model extension. 

The EATX metric described in Section~\ref{sec:anypath_cost} was originally designed considering that nodes transmit at a single bit rate. To account for multiple bit rates, we introduce the expected anypath transmission time (EATT) metric. For EATT, the hyperlink cost $d_{iJ}^{\,(r)}$ for each rate $r \in R$ is defined as

\vspace{-0.05in}
\begin{equation}
\label{eq:d_iJ_ETT} d_{iJ}^{\,(r)} = \frac{1}{p_{iJ}^{(r)}} \times \frac{s}{r},
\vspace{-0.05in}
\end{equation}
where $p_{iJ}^{(r)}$ is the hyperlink delivery ratio defined in~(\ref{eq:p_iJ}), $s$ is the maximum packet size, and $r$ is the bit rate. The cost $d_{iJ}^{\,(r)}$ is the time taken to transmit a packet of size~$s$ at a bit rate~$r$ over a lossy hyperlink with delivery ratio~$p_{iJ}^{(r)}$. The EATT metric is a generalization of the expected transmission time (ETT) metric~\cite{draves04b}, commonly used in single-path wireless routing. Note that for each bit rate $r \in R$, we have a different delivery ratio~$p_{iJ}^{(r)}$, which usually decreases for higher rates. This behavior imposes a tradeoff; a higher bit rate decreases the time of a single packet transmission (i.e., $s/r$ decreases), but it usually increases the number of transmissions required for a packet to be successfully received (i.e., $1/p_{iJ}^{(r)}$ increases).

The remaining anypath cost~$D_J^{(r)}$ now also depends on the transmission rate, since the delivery ratios change for each rate. Since both the hyperlink cost and the remaining anypath cost depend on the bit rate, node~$i$ has a different anypath cost~$D_i^{(r)} = d_{iJ}^{\,(r)} + D_J^{(r)}$ for each forwarding set~$J$ and for each transmission rate $r \in R$. The remaining anypath cost~$D_J^{(r)}$ for a rate $r \in R$ is defined as
\begin{equation}
\label{eq:D_J_r_def} D_J^{(r)} = \sum_{j \in J} w_{ij}^{(r)}D_j, \mathrm{\ with\ }\sum_{j \in J} w_{ij}^{(r)} = 1,
\end{equation}
where the weight~$w_{ij}^{(r)}$ in~(\ref{eq:D_J_r_def}) is the probability of node~$j$ being the relaying node for node~$i$, and $D_j = \min_{r \in R} D_j^{(r)}$ is the lowest cost from node~$j$ to the destination among all rates. 
The weight~$w_{ij}^{(r)}$ is then defined as
\begin{equation}
w_{ij}^{(r)} = \frac{P[\,X_1^{(r)}=0,\ldots,X_{j-1}^{(r)}=0,X_j^{(r)}=1\,]}{1-P[\,X_1^{(r)}=0,X_2^{(r)}=0,\ldots,X_n^{(r)}=0\,]},
\end{equation}
where $X_j^{(r)}$ is a random variable equal to 0 or 1 if $j \in J$ respectively loses or receives a packet transmitted at rate~$r$ by node~$i$. For the case of independent losses, the weight $w_{ij}^{(r)}$ can be simplified to
\vspace{-0.1in}
\begin{equation}
w_{ij}^{(r)} = \frac{\displaystyle p_{ij}^{(r)}\prod_{k=1}^{j-1}\left[1-p_{ik}^{(r)}\right]}{1-\displaystyle\prod_{j \in J} \left[1-p_{ij}^{(r)}\right]}.
\end{equation}

We address the problem of finding both the forwarding set and the transmission rate that minimize the overall cost to reach a particular destination. We call this the {\it shortest multirate anypath problem}, which generalizes the shortest-anypath problem~\cite{dubois-ferriere10} for the multirate scenario. Interestingly, the shortest multirate anypath will always have equal or lower cost than a shortest anypath using a single transmission rate. Among all possible multirate anypaths between two nodes, we also have the shortest single-rate anypaths. As a result, the cost of the shortest multirate anypath can never be higher than the cost of any of the shortest single-rate anypaths.

\section{Finding the Shortest Multirate Anypath}
\label{sec:shortest}

In this section we introduce our shortest-anypath algorithms. In Section~\ref{sec:single-rate}, we present the Shortest Anypath First (SAF) and the Anypath Bellman-Ford (ABF) algorithms used in a single-rate network with the EATX metric. Similar single-rate algorithms were also proposed by Chachulski~\cite{chachulski07b}. We, however, derived these algorithms independently and later in Section~\ref{sec:multi-rate} we introduce a generalization of these algorithm for multiple rates. Surprisingly, the Shortest Multirate Anypath First (SMAF) and the Multirate Anypath Bellman-Ford (MABF) algorithms have roughly the {\it same} running time as the corresponding shortest single-path algorithms for multirate.
We only show the proof of optimality of the multirate algorithms, since by definition this also implies the optimality of the single-rate algorithms.

\vspace{-.1in}
\subsection{The Single-Rate Case}
\label{sec:single-rate}

We now present the Shortest Anypath First (SAF) algorithm used in the simpler single-rate scenario. Given a graph $G=(V,E)$, the algorithm calculates the shortest anypaths from all nodes to a destination~$d$. For every node $i \in V$ we keep an estimate~$D_i$, which is an upper-bound on the cost of the shortest anypath from~$i$ to~$d$. In addition, we also keep a forwarding set $F_i$ for every node, which stores the set of nodes used as the next hops to reach~$d$. Finally, we keep two data structures, namely~$S$ and~$Q$. The $S$ set stores the set of nodes for which we already have a shortest anypath defined. 
We store each node $i \in V - S$ for which we still do not have a shortest anypath in a priority queue~$Q$ keyed by their $D_i$ values.

Lines~1--3 initialize the state variables $D_i$ and $F_i$ and line~4 sets to zero the cost from node $d$ to itself. Lines~5--6 initialize the $S$ and $Q$ data structures. Initially, we do not have the shortest anypath from any node, so $S$ is initially empty and thus $Q$ contains all vertices. 
As in the shortest-path algorithm, the Shortest Anypath First algorithm is composed of $|V|$
rounds, dictated by the number of elements initially in~$Q$. At each round, the {\sc Extract-Min} procedure extracts the node with the minimum cost to~$d$ from~$Q$. Let this node be $j$. At this point, $j$ is settled and inserted into~$S$, since the shortest anypath from~$j$ to the destination is now known. 
For each incoming edge $(i,j) \in E$, we check if the cost~$D_i$ is larger \hfill than \hfill the \hfill cost \hfill $D_j$. \hfill If \hfill that \hfill is \hfill the \hfill case, \hfill then \hfill node~$j$ \hfill is \hfill added \hfill to

\begin{algorithm}{Shortest-Anypath-First}{G,d}
\begin{FOR}{\EACH \text{\bf node} i \IN V}
D_i \= \infty \\
F_i \= \emptyset
\end{FOR} \\
D_d \= 0 \\
S \= \emptyset \\
Q \= V \\
\begin{WHILE}{Q \neq \emptyset}
j \= \CALL{Extract-Min}(Q)\\
S \= S \cup \{j\} \\
\begin{FOR}{\EACH \text{\bf incoming edge} (i,j) \IN E}
J \= F_i \cup \{j\} \\
\begin{IF}{D_i > D_j}
D_i \= d_{iJ} + D_J \\
F_i \= J
\end{IF}
\end{FOR} 
\end{WHILE}
\end{algorithm}

\noindent  the forwarding set $F_i$ and the cost $D_i$ is updated.

Fig.~\ref{fig:saf-execution} shows the step-by-step execution of the SAF algorithm using the EATX metric. The link weights represent the link delivery ratios. Fig.~\ref{fig:saf-execution-a} depicts the graph right after the initialization and Fig.~\ref{fig:saf-execution-b}--\ref{fig:saf-execution-h} show each iteration of the algorithm. The value inside a node~$i$ presents its cost $D_i$ to the destination~$d$ (i.e., the end-to-end expected number of transmissions), and the arrows in boldface present the shortest anypath to~$d$. Nodes with two circles are the settled nodes in~$S$; each iteration settles a new node. The graph in Fig.~\ref{fig:saf-execution-h} shows the result of the SAF algorithm right after settling the last node.

The running time of the SAF algorithm depends on how~$Q$ is implemented. Assuming we have a Fibonacci heap, the cost of each of the $|V|$ {\sc Extract-Min} operations in line~8 takes $O(\log V)$, with a total aggregated time of $O(V \log V)$. 
The node cost in line~13 can be updated in a constant time, as follows. Let $D_i$ be the cost using a forwarding set~$J$ and let~$D_i'$ be the cost using $J' = J \cup \{j\}$, with $D_j \geq D_k$ for all $k \in J$. The new cost $D_i'$ can then be calculated from the previous cost $D_i$ as 
\setlength{\arraycolsep}{0.0em}%
\begin{eqnarray}
\nonumber D_i' &{}={}& d_{iJ'} + D_{J'} \\
\nonumber      &{}={}& \frac{p_{iJ}}{p_{iJ'}}d_{iJ} + \frac{p_{iJ}}{p_{iJ'}}D_J + \left(1-\frac{p_{iJ}}{p_{iJ'}}\right)D_j \\
               &{}={}& \frac{p_{iJ}}{p_{iJ'}}D_i + \left(1-\frac{p_{iJ}}{p_{iJ'}}\right)D_j,
\end{eqnarray}
\setlength{\arraycolsep}{5pt}%
where $p_{iJ}/p_{iJ'}$ scales $d_{iJ}$ and $D_J$ to account for the new forwarding set~$J'$ while $1-p_{iJ}/p_{iJ'}$ is the weight of~$D_j$ in~$D_{J'}$. The cost~$D_i$ in line~13 is then updated using just the previous cost and~$D_j$. The {\bf for} loop of lines 10--13 takes $O(E)$ aggregated time and the total complexity is $O(V\log V + E)$, which is the same complexity as Dijkstra's algorithm.

We also present the Anypath Bellman-Ford (ABF) algorithm, which can be implemented in a distributed fashion. The ABF algorithm reduces the complexity of the Bellman-Ford anypath generalization proposed in~\cite{dubois-ferriere10} from exponential to polynomial time due to a property we show in Section~\ref{sec:multi-rate}. 
As in the regular Bellman-Ford, the ABF algorithm is also composed of at most $|V|-1$ rounds. At each round, every node~$i$ stores its neighbors in a priority queue~$Q$ keyed by their cost.  
We then check each neighbor~$j$ in ascending order of cost~$D_j$, and verify whether $D_i$ is larger than~$D_j$. If that is the case, node $i$ includes $j$ in its forwarding set and updates its distance accordingly. Intuitively, the algorithm works in the same expanding-ring fashion as the regular Bellman-Ford, settling at each round the costs of the nodes one hop further away from the destination. Since an anypath can not be longer than $|V|-1$ hops, the algorithm converges after at most $|V|-1$ iterations.
\\
\begin{algorithm}{Anypath-Bellman-Ford}{G,d}
\begin{FOR}{\EACH \text{\bf node} i \IN V}
D_i \= \infty \\
F_i \= \emptyset
\end{FOR} \\
D_d \= 0 \\
\begin{FOR}{t \= 1 \TO |V|-1}
\begin{FOR}{\EACH \text{\bf node} i \IN V}
J \= \emptyset \\
Q \= \CALL{Get-Neighbors}(i) \\
\begin{WHILE}{Q \neq \emptyset}
j \= \CALL{Extract-Min}(Q)\\
J \= J \cup \{j\} \\
\begin{IF}{D_i > D_j}
D_i \= d_{iJ} + D_J \\
F_i \= J
\end{IF}
\end{WHILE}
\end{FOR}
\end{FOR}
\end{algorithm}

\begin{figure}[ht!]
\centering
\subfigure[]{
  \label{fig:abf-execution-a}
  \includegraphics[width=.22\textwidth]{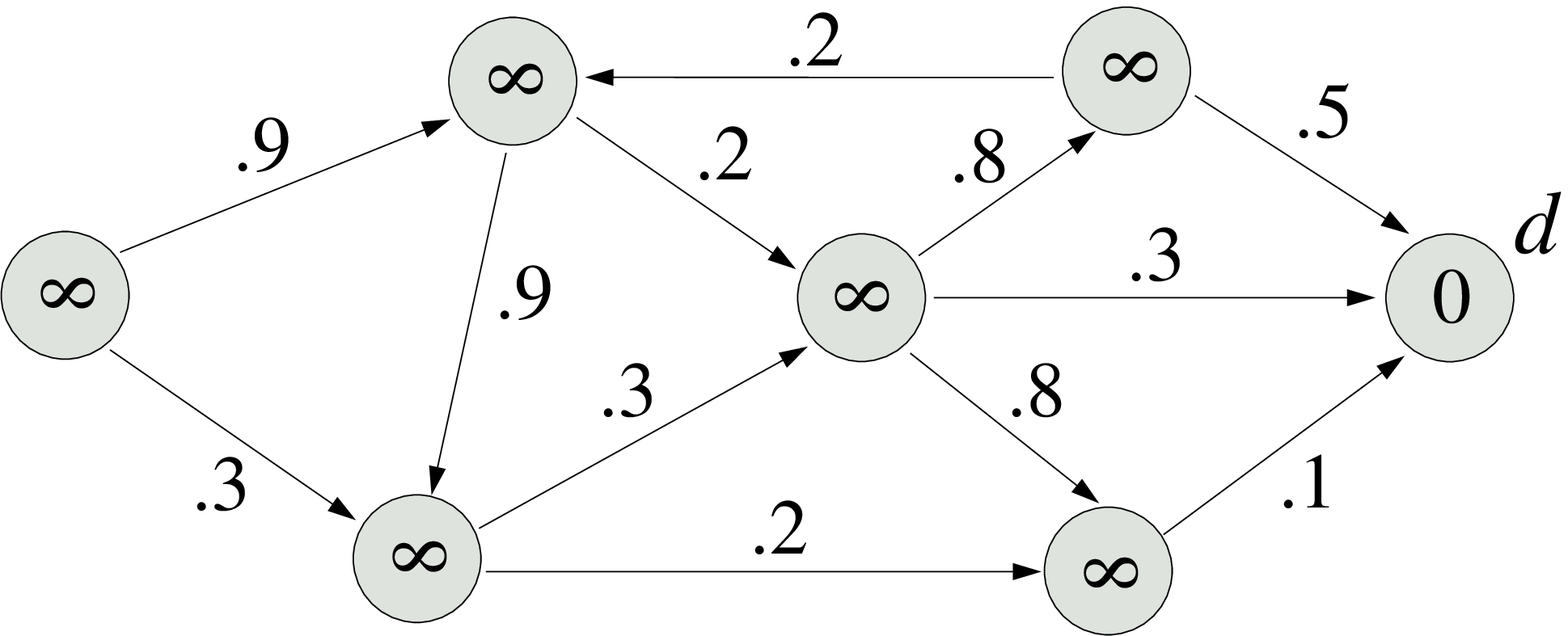}
}
\subfigure[]{
  \label{fig:abf-execution-b}
  \includegraphics[width=.22\textwidth]{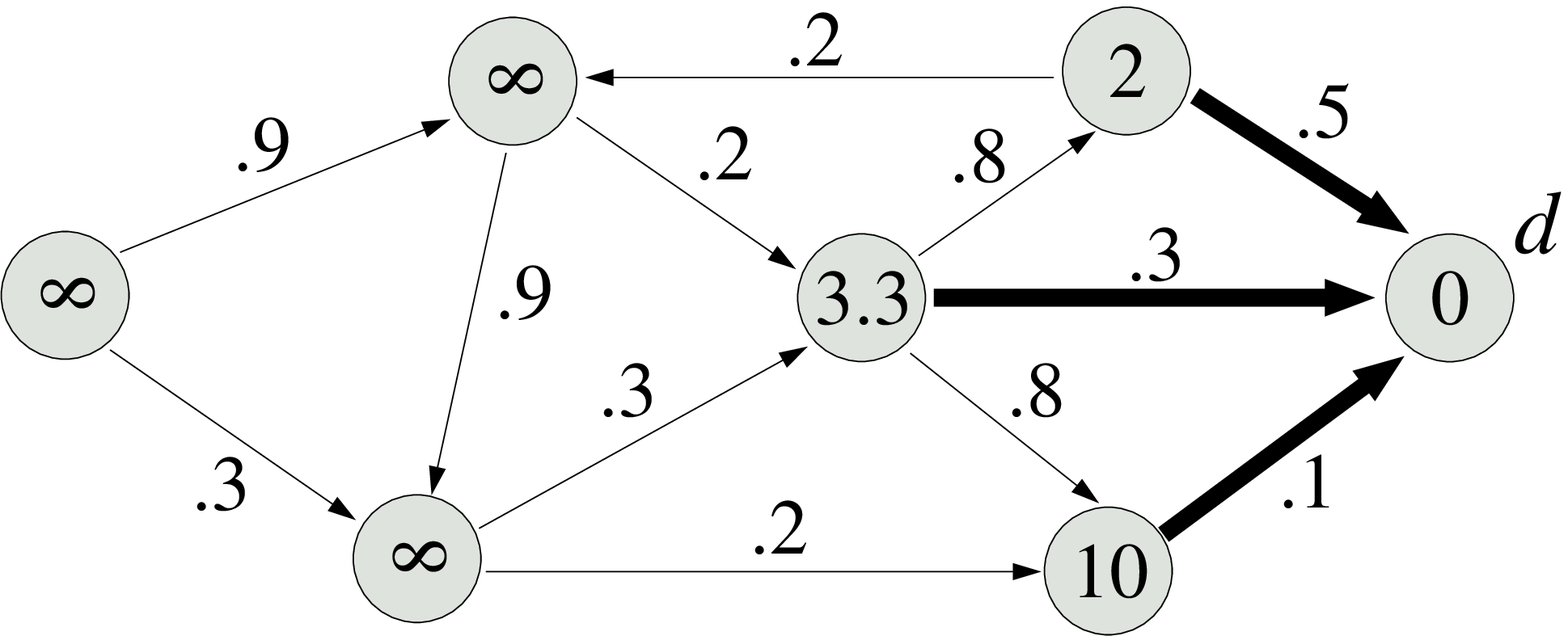}
}
\subfigure[]{
  \label{fig:abf-execution-c}
  \includegraphics[width=.22\textwidth]{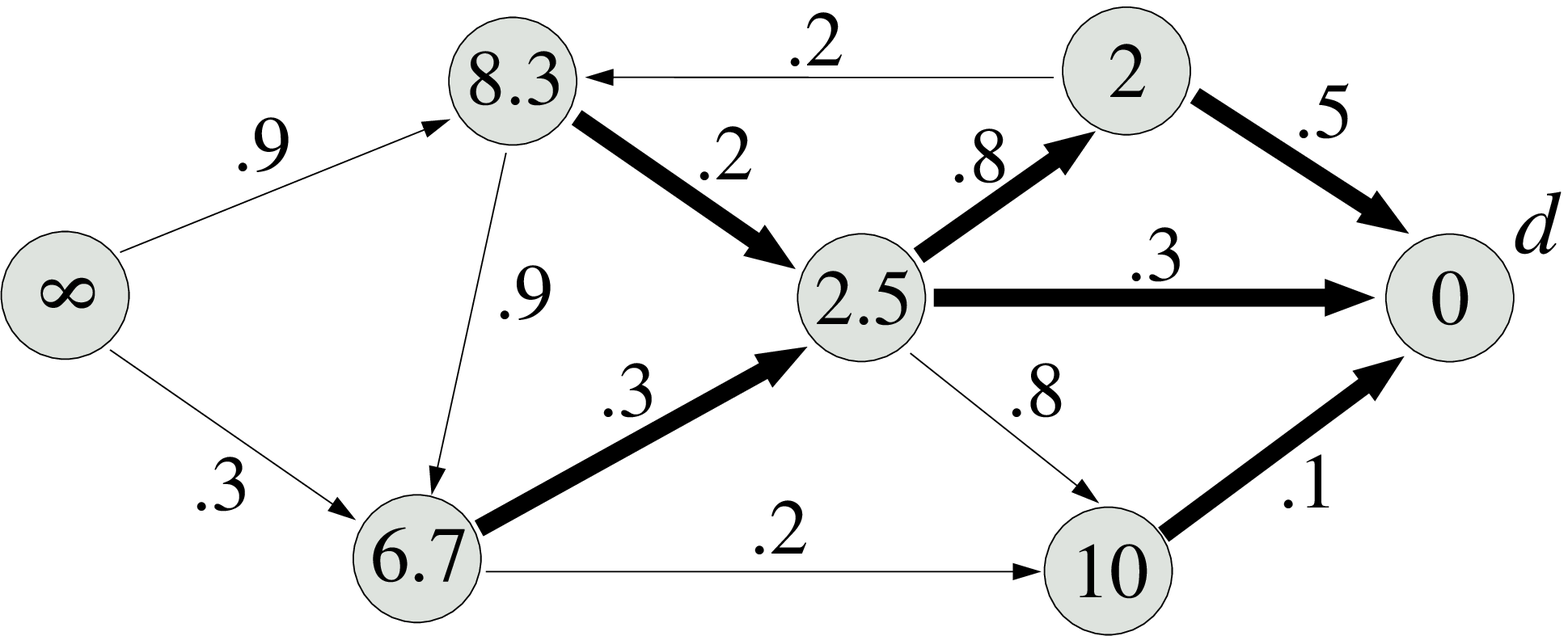}
}
\subfigure[]{
  \label{fig:abf-execution-d}
  \includegraphics[width=.22\textwidth]{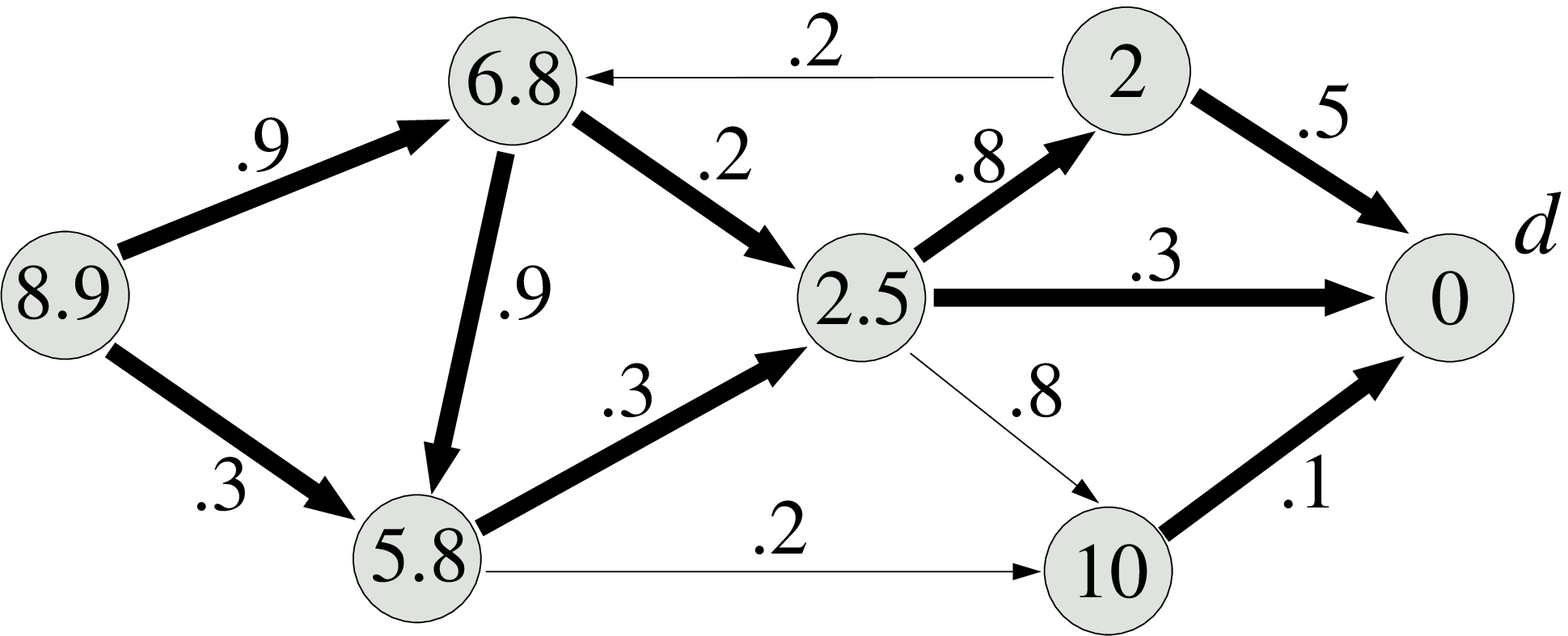}
}
\subfigure[]{
  \label{fig:abf-execution-e}
  \includegraphics[width=.22\textwidth]{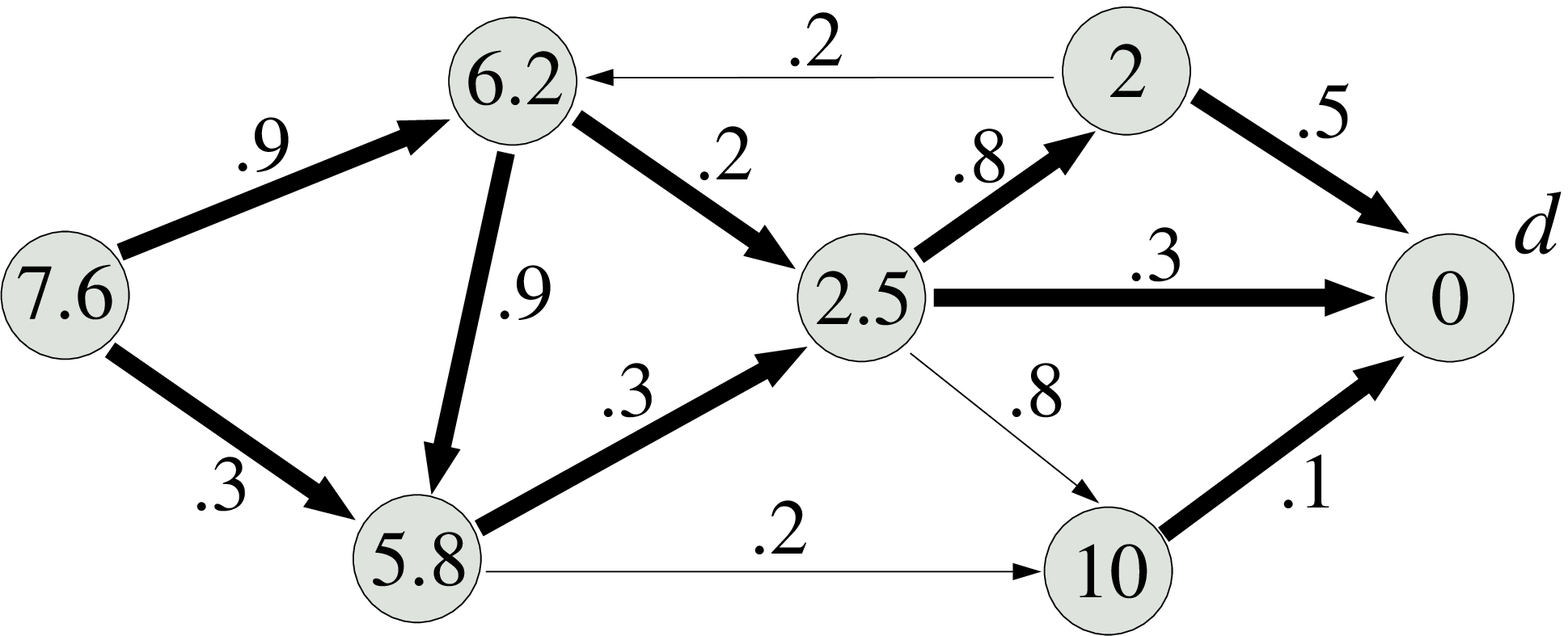}
}
\subfigure[]{
  \label{fig:abf-execution-f}
  \includegraphics[width=.22\textwidth]{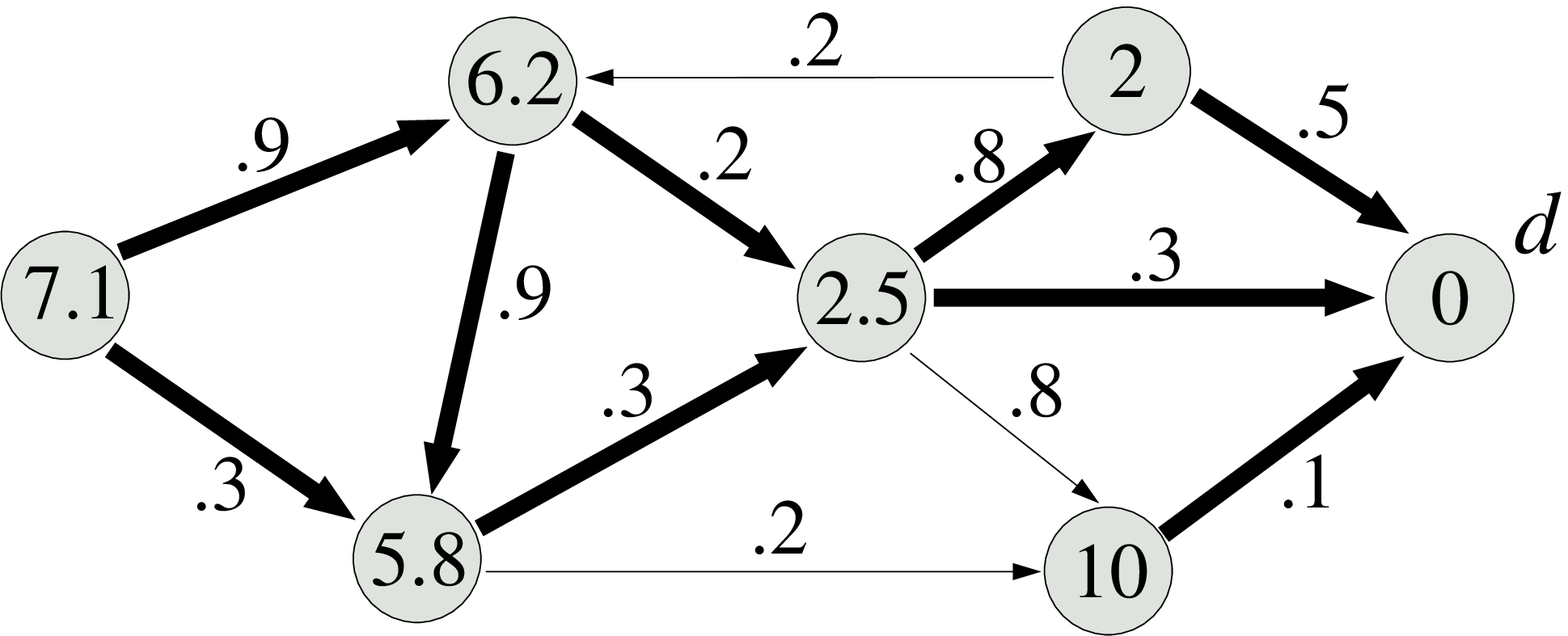}
}
\caption{Execution of the Anypath Bellman-Ford (ABF) algorithm from every node to~$d$. The weights represent the link delivery ratios, the value inside a node~$i$ represents its cost $D_i$ to~$d$ (i.e., the end-to-end expected number of transmissions), and the arrows in boldface represent the shortest anypath to~$d$. (a) The situation just after the initialization. (b)--(f) The situation after each successive iteration. Node costs are updated considering the costs in the previous iteration.}
\label{fig:abf-execution}
\vspace{-0.3in}
\end{figure}

Fig.~\ref{fig:abf-execution} shows the step-by-step execution of the ABF algorithm using the EATX metric. Fig.~\ref{fig:abf-execution-a} depicts the graph just after the initialization and Fig.~\ref{fig:abf-execution-b}--\ref{fig:abf-execution-f} show each iteration of the algorithm. The value inside a node~$i$ presents its cost $D_i$ to~$d$ (i.e., the end-to-end expected number of transmissions), and the arrows in boldface present the shortes anypath to~$d$. The graph in Fig.~\ref{fig:abf-execution-f} shows the result of the ABF algorithm. 

The running time of the ABF algorithm depends on how~$Q$ is implemented. Assuming a Fibonacci heap, each of the {\sc Extract-Min} operations in line~10 takes $\log(V)$ at the most. The {\bf for} loop in lines~6--14 runs once for each link, for a total aggregated time of $O(E\log V)$. The total complexity of the ABF algorithm is then $O(VE\log V)$, which is only a factor of $\log V$ higher than the regular Bellman-Ford algorithm.

\subsection{The Multirate Case}
\label{sec:multi-rate}

\begin{figure*}[ht!]
\centering
\subfigure[]{
  \label{fig:multirate-result-a}
  \includegraphics[width=.22\textwidth]{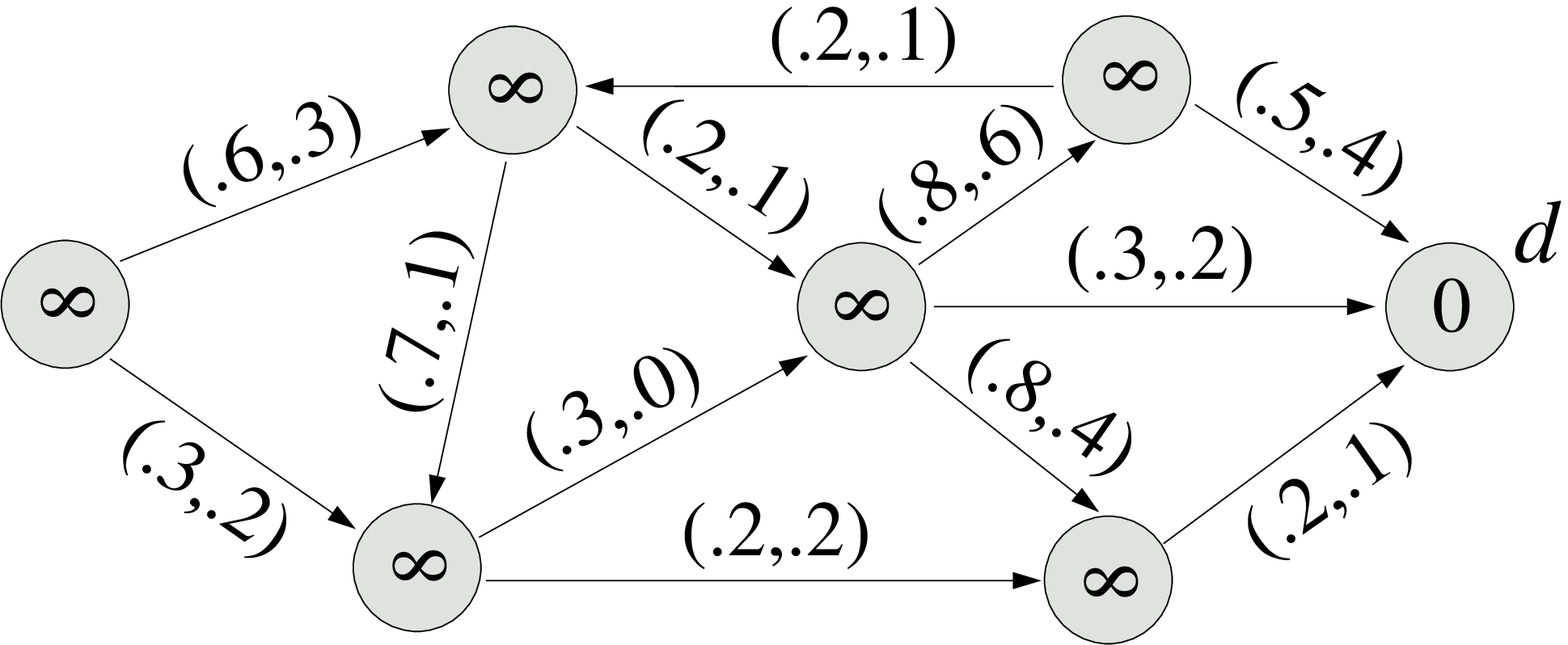}
}
\subfigure[]{
  \label{fig:multirate-result-b}
  \includegraphics[width=.22\textwidth]{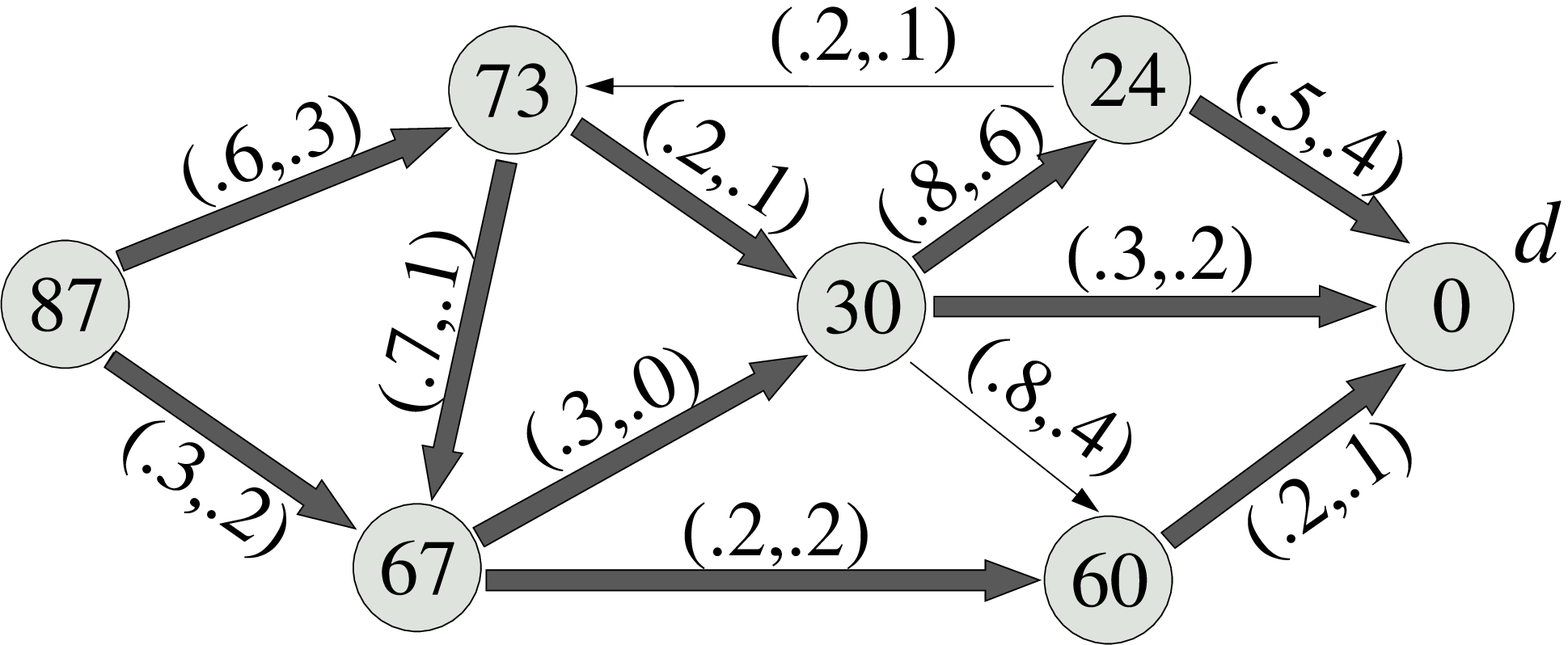}
}
\subfigure[]{
  \label{fig:multirate-result-c}
  \includegraphics[width=.22\textwidth]{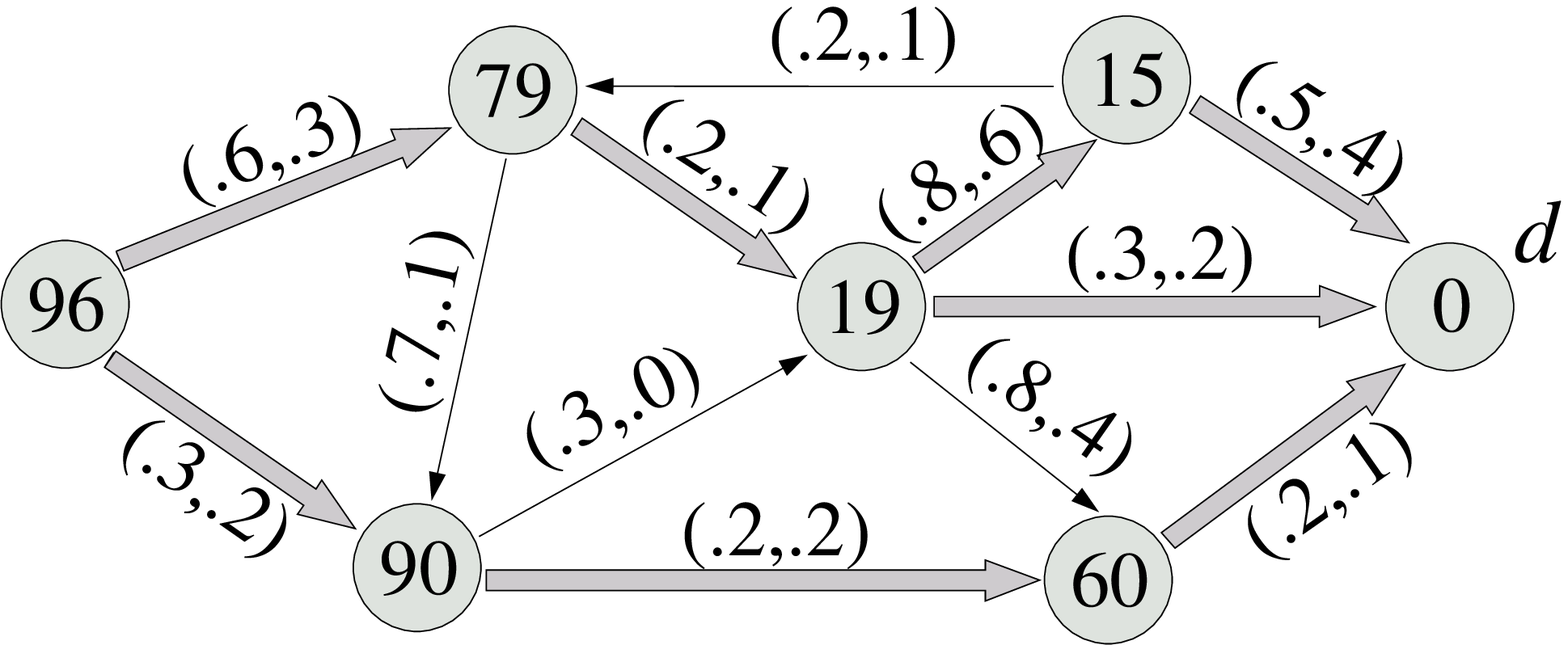}
}
\subfigure[]{
  \label{fig:multirate-result-d}
  \includegraphics[width=.22\textwidth]{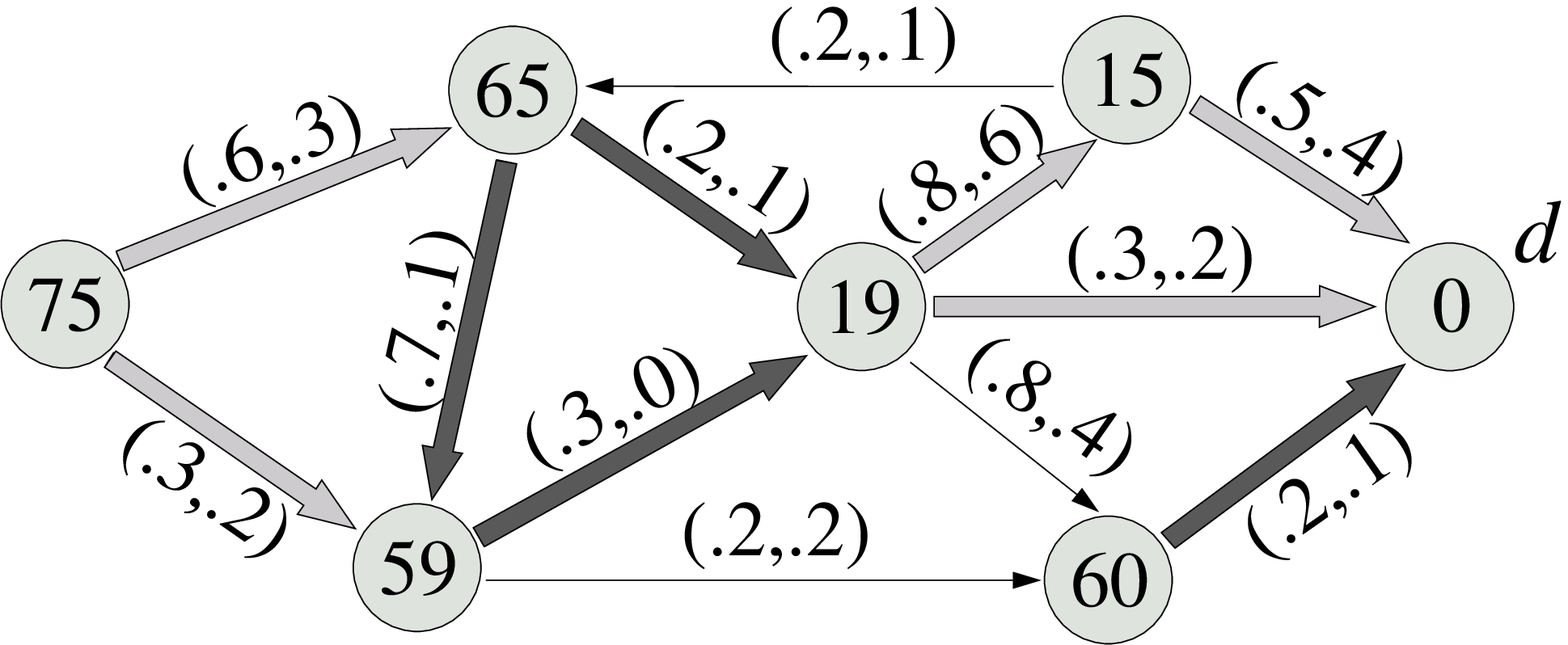}
}
\caption{The advantage of multirate over single-rate anypath routing. (a) Each pair of weights $(w_1,w_2)$ represents the link delivery ratio at 1~Mbps and 2~Mbps, respectively. The value inside a node~$i$ is its cost $D_i$ to~$d$ using the EATT metric (i.e., the end-to-end expected transmissions time in milliseconds), and the arrows in boldface represent the shortest multirate anypath to~$d$. (b) The shortest anypath at 1~Mbps. (c) The shortest anypath at 2~Mbps. (d) The shortest multirate anypath using both 1~Mbps (dark gray arrows) and 2~Mbps (light gray arrows) rates. The multirate anypath in (d) takes advantage of the best rate at each node to yield the lowest cost.}
\label{fig:multirate-result}
\vspace{-0.2in}
\end{figure*}

We now generalize the SAF algorithm to support multiple transmission rates, introducing the Shortest Multirate Anypath First (SMAF) algorithm. 
For each node $i \in V$, we now keep a different cost estimate~$D_i^{(r)}$ for every rate $r \in R$. The estimate~$D_i^{(r)}$ is an upper-bound on the cost of the shortest anypath from~$i$ to~$d$ using transmission rate~$r$. In addition, we also keep its corresponding forwarding set~$F_i^{(r)}$, which stores the set of next hops used for~$i$ to reach~$d$ using~$r$. We use~$D_i$ and~$F_i$ without the indicated rates to store the minimum cost estimate among all rates and its corresponding forwarding set, respectively. We also keep a transmission rate $T_i$ for every node, which stores the optimal rate used to reach~$d$. 
\\
\begin{algorithm}{Shortest-Multirate-Anypath-First}{G,d}
\begin{FOR}{\EACH \text{\bf node} i \IN V}
D_i \= \infty \\
F_i \= \emptyset \\
T_i \= \NIL \\
\begin{FOR}{\EACH \text{\bf rate} r \IN R}
D_i^{\,(r)} \= \infty \\
F_i^{\,(r)} \= \emptyset
\end{FOR}
\end{FOR} \\
D_d \= 0 \\
S \= \emptyset \\
Q \= V \\
\begin{WHILE}{Q \neq \emptyset}
j \= \CALL{Extract-Min}(Q)\\
S \= S \cup \{j\} \\
\begin{FOR}{\EACH \text{\bf incoming edge} (i,j) \IN E}
\begin{FOR}{\EACH \text{\bf rate} r \IN R}
J \= F_i^{(r)} \cup \{j\} \\
\begin{IF}{D_i^{(r)} > D_j}
D_i^{\,(r)} \= d_{iJ}^{\,(r)} + D_J^{(r)} \\
F_i^{\,(r)} \= J \\
\begin{IF}{D_i > D_i^{(r)}}
D_i \= D_i^{(r)} \\
F_i \= F_i^{(r)} \\
T_i \= r
\end{IF}
\end{IF}
\end{FOR} 
\end{FOR}
\end{WHILE}
\end{algorithm}
The idea of the SMAF algorithm is that each node $i \in V$ has an independent cost estimate~$D_i^{(r)}$ for each rate $r \in R$ and we keep the minimum of these estimates as the node cost~$D_i$. At each round of the {\bf while} loop, the node with the minimum cost from~$Q$ is settled. Let this node be~$j$. For each incoming edge $(i,j) \in E$, we check for every rate $r \in R$ if the cost $D_i^{(r)}$ is larger than the cost $D_j$ of the node just settled. If this is the case, then node~$j$ is added to the forwarding set $F_i^{(r)}$ of that specific rate and cost $D_i^{(r)}$ is updated accordingly. If the new cost~$D_i^{(r)}$ is lower than the node cost~$D_i$, we update~$D_i$ as well as the forwarding set~$F_i$ and transmission rate~$T_i$ to reflect the new minimum.
The running time of the Shortest Multirate Anypath First algorithm is $O(V \log V + ER)$, which is the same running time of the shortest single-path algorithm for multiple rates.

\vfill
We also introduce the Multirate Anypath Bellman-Ford (MABF), a generalization of the ABF algorithm for multiple transmission rates. The MABF uses the same idea of keeping a different estimate cost~$D_i^{(r)}$ for each rate $r \in R$ and taking the minimum as the node cost. The running time of the MABF algorithm is $O(VE\log V + VER)$, where $O(VE\log V)$ is the aggregated time of the {\sc Extract-Min} operations and $O(VER)$ is the aggregated time of the {\bf for} loop in lines 14--22.
\vfill
\begin{algorithm}{Multirate-Anypath-Bellman-Ford}{G,d}
\begin{FOR}{\EACH \text{\bf node} i \IN V}
D_i \= \infty \\
F_i \= \emptyset \\
T_i \= \NIL \\
\begin{FOR}{\EACH \text{\bf rate} r \IN R}
D_i^{\,(r)} \= \infty \\
F_i^{\,(r)} \= \emptyset
\end{FOR}
\end{FOR} \\
D_d \= 0 \\
\begin{FOR}{t \= 1 \TO |V|-1}
\begin{FOR}{\EACH \text{\bf node} i \IN V}
Q \= \CALL{Get-Neighbors}(i) \\
\begin{WHILE}{Q \neq \emptyset}
j \= \CALL{Extract-Min}(Q)\\
\begin{FOR}{\EACH \text{\bf rate} r \IN R}
J \= F_i^{(r)} \cup \{j\} \\
\begin{IF}{D_i^{(r)} > D_j}
D_i^{(r)} \= d_{iJ} + D_J \\
F_i^{(r)} \= J\\
\begin{IF}{D_i > D_i^{(r)}}
D_i \= D_i^{(r)} \\
F_i \= F_i^{(r)} \\
T_i \= r
\end{IF}
\end{IF}
\end{FOR}
\end{WHILE}
\end{FOR}
\end{FOR}
\end{algorithm}
Fig.~\ref{fig:multirate-result} shows the advantage of multirate over single-rate anypath routing. Fig.~\ref{fig:multirate-result-a} depicts a network topology, with each pair of weights $(w_1,w_2)$ representing the link delivery ratio at 1~Mbps and 2~Mbps, respectively. The value inside each node~$i$ is its cost~$D_i$ to~$d$ using the EATT metric (i.e., the end-to-end expected transmission time in milliseconds). Fig.~\ref{fig:multirate-result-b} shows the shortest single-rate anypath at 1~Mbps (i.e., cost of 87~ms); Fig.~\ref{fig:multirate-result-c} shows the same result at 2~Mbps (i.e., cost of 96~ms). In our example, nodes further from~$d$ have a lower cost when transmitting at 1~Mbps, as shown in~(b). On the other hand, nodes closer to~$d$ achieve a lower cost at 2~Mbps, as shown in~(c). Using a single transmission rate therefore does not allow nodes to reduce their costs by much. The shortest multirate anypath in~\ref{fig:multirate-result-d}, however, takes advantage of the best rate at each node to yield the lowest cost (i.e., 75~ms).

\subsection{Optimality}

In order to prove the optimality of the algorithm, we first introduce five lemmas that show a few properties of multirate anypath routing. We use~$\delta_i^{(r)}$ as the cost of the shortest multirate anypath from a node~$i$ to the destination~$d$, when~$i$ transmits at a fixed rate $r \in R$. Likewise, $\phi_i^{(r)}$ represents the corresponding forwarding set used in this multirate anypath. We use~$\delta_i$ without the indicated rate to represent the cost of the shortest multirate anypath from~$i$ to~$d$ via the optimal forwarding set~$\phi_i$ and optimal transmission rate $\rho \in R$. That~is, $\delta_i = \min_{r \in R} \delta_i^{(r)}$, $\rho = \argmin_{r \in R} \delta_i^{(r)}$, and $\phi_i = \phi_i^{(\rho)}$. We use~$D_i$ as the cost of a particular multirate anypath from~$i$ to~$d$, but not necessarily the shortest one. 
The proof for each of these lemmas is available in the Appendix.

\begin{lem}
\label{lem:comparison} 
{\it For a fixed transmission rate, let $D_i$ be the cost of a node~$i$ via forwarding set~$J$ and let $D_i'$ be the cost via forwarding set $J' = J \cup \{k\}$, where $D_k \geq D_j$ for every node $j \in J$. We have $D_i' \leq D_i$ if and only if $D_i \geq D_k$.}
\end{lem}

We use Lemma~\ref{lem:comparison} for the comparisons in line~12 of the SAF and ABF algorithms, as well as in line~17 of the SMAF and line~16 of the MABF algorithms. By this lemma, if the cost~$D_i$ via~$J$ is larger than the cost $D_k$ of a neighbor node~$k$, with $D_k \geq D_j$ for all $j \in J$, then the cost~$D_i'$ via $J'= J \cup \{k\}$ is always lower than $D_i$; that is, it is always beneficial to include node~$k$ in the forwarding set in order to obtain a lower cost to the destination.

\begin{lem}
\label{lem:acyclic} 
{\it The lowest cost~$\delta_i$ of a node~$i$ is always larger than or equal to the lowest cost $\delta_j$ of any node~$j$ in the optimal forwarding set $\phi_i$. That is, we have $\delta_i \geq \delta_j$ for all $j \in \phi_i$.}
\end{lem}

Lemma~\ref{lem:acyclic} guarantees that if a node~$i$ uses another node~$j$ in its optimal forwarding set~$\phi_i$, then cost~$\delta_i$ can never be smaller than~$\delta_j$. This is equivalent to the restriction that all weights in the graph must be nonnegative. Note that this property also implies that no cycles can occur in an anypath. If a node~$i$ uses a node~$j$ in its forwarding set, then according to Lemma~\ref{lem:acyclic} we must have $\delta_i \geq \delta_j$. If, however, node~$j$ also uses node~$i$ in its forwarding set (a cycle), we must also have $\delta_i \leq \delta_j$. Therefore, we can only have a cycle when $\delta_i = \delta_j$. However, we show in the next lemma that these cycles do not change the anypath cost. As a result, we use strict comparisons in our algorithms to prevent such cycles from occuring in practice. Due to the same reason, larger cycles with more than two nodes can not occur either and the shortest anypath is always a directed acyclic graph (DAG).

\begin{lem}
\label{lem:extra} 
{\it For any transmission rate, if a node~$i$ uses a node~$k$ in its optimal forwarding set~$\phi_i$ and $\delta_i = \delta_k$, we can safely remove~$k$ from~$\phi_i$ without changing~$\delta_i$. The link $(i,k)$ is said to be ``redundant.''}
\end{lem}

By Lemma~\ref{lem:extra}, if the costs $\delta_i = \delta_k$ of two nodes~$i$ and~$k$ are the same, then the cost~$\delta_i$ via forwarding set~$\phi_i$ is the same as the cost via the set~$\phi_i - \{k\}$. That is, the cost of node~$i$ does not change if it uses~$k$ in its forwarding set or not.

\begin{lem}
\label{lem:order} 
{\it If the lowest costs from the neighbors of a node~$i$ to a given destination are $\delta_1 \leq \delta_2 \leq \ldots \leq \delta_n$, then the optimal forwarding set~$\phi_i^{(r)}$ is always of the form $\phi_i^{(r)} = \{1,2,\ldots,k\}$, for some $k \in \{1,2,\ldots,n\}$.}
\end{lem}

According to Lemma~\ref{lem:order}, the best forwarding set~$\phi_i^{(r)}$ for transmission rate $r \in R$ is a subset of neighbors with the lowest costs to the destination. That is, given a set of neighbors with costs $\delta_1 \leq \delta_2 \leq \ldots \leq \delta_n$, the best forwarding set~$\phi_i^{(r)}$ when using rate $r \in R$ is always one of $\{1\}$, $\{1,2\}$, $\{1,2,3\},\ldots,\{1,2,\ldots,n\}$. As a result, forwarding sets with gaps between the neighbors, such as $\{2,3\}$ or $\{1,4\}$, can never yield the lowest cost to the destination. This property is the key factor that allows us to reduce the complexity of the proposed algorithms from exponential to polynomial time. For $n$ neighbors, we do not have to test every one of the $2^n-1$ possible forwarding sets. Instead, we only need to check at most $n$ forwarding sets.

\begin{lem} 
\label{lem:fset} 
{\it For a given transmission rate $r \in R$, assume that $\phi_i^{(r)} = \{1,2,\ldots,k\}$ with costs $\delta_1 \leq \delta_2 \leq \ldots \leq \delta_k$. If $D_i^j$ is the cost from node~$i$ using transmission rate~$r$ via forwarding set $\{1,2,\ldots,j\}$, for $1 \leq j \leq k$, then we always have $D_i^1 \geq D_i^2 \geq \ldots \geq D_i^k = \delta_i^{(r)}$.}
\end{lem}

Lemma~\ref{lem:fset} explains another important property necessary for the proposed routing algorithms to converge. Assuming that the best forwarding set $\phi_i^{(r)}$ for transmission rate $r \in R$ is $\phi_i^{(r)} = \{1,2,\ldots,k\}$ with costs $\delta_1 \leq \delta_2 \leq \ldots \leq \delta_k$, the cost $D_i$ monotonically decreases as we use each of the forwarding sets $\{1\},\{1,2\},\{1,2,3\},\ldots,\{1,2,\ldots,k\}$.

We now present the proof of optimality of our algorithms. 
\begin{thm}
\label{thm:saf_correctness}
{\it
Optimality of the SMAF algorithm.

Let $G = (V,E)$ be a weighted directed graph and let $d$ be the destination. After running the Shortest Multirate Anypath First algorithm on $G$, we have $D_i = \delta_i$ for all nodes $i \in V$.}
\end{thm}
\begin{IEEEproof}
This proof is similar to the proof of Dijkstra's algorithm~\cite{cormen01}. We show that for each node $s \in V$, we have $D_s = \delta_s$ at the time $s$ is added to $S$.

For the purpose of contradiction, let $s$ be the first node added to $S$ for which $D_s \neq \delta_s$. We must have $s \neq d$ because~$d$ is the first node added to $S$ and $D_d = \delta_d = 0$ at that time. Just before adding $s$ to $S$, we also have that $S$ is not empty, since $s \neq d$ and $S$ must contain at least $d$. We assume that there must be a multirate anypath from $s$ to $d$, otherwise $D_s = \delta_s = \infty$, which contradicts our initial assumption that $D_s \neq \delta_s$. If there is at least one multirate anypath, there is a shortest multirate anypath~$\alpha$ from $s$ to $d$. Let us consider a cut $(V - S, S)$ of~$\alpha$, such that we have $s \in V - S$ and $d \in S$, as shown in Fig.~\ref{fig:saf_proof}. Let the set~$J$ be composed of nodes in $V - S$ that have an outgoing link to a node in~$S$. 
Likewise, let the set~$K$ be composed of nodes in~$S$ that have an incoming link from a node in~$V - S$.

\begin{figure}[h!]
\centering
\includegraphics[width=.40\textwidth]{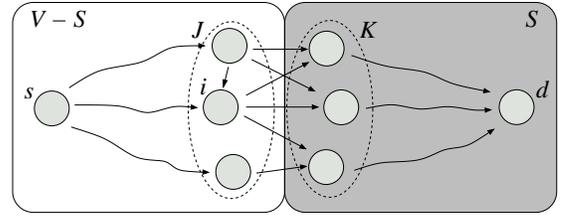}
\caption{The shortest multirate anypath~$\alpha$ from~$s$ to~$d$. Set~$S$ must be nonempty before node~$s$ is inserted into it, since it must contain at least~$d$. We consider a cut $(V - S, S)$ of $\alpha$, such that we have $s \in V - S$ and $d \in S$. Nodes~$s$ and~$d$ are distinct but we may have no hyperlinks between $s$ and $J$, such that $J = \{s\}$, and also between $K$ and $d$, such that $K = \{d\}$.}
\label{fig:saf_proof}
\end{figure}

Without loss of generality, assume that node $i \in J$ has the lowest cost to~$d$ among all nodes in~$V - S$. That is, $\delta_i \leq \delta_j$ for all $j \in V - S$. We claim that every edge leaving node~$i$ must necessarily cross the cut $(V - S, S)$. Thus, for every edge~$(i,j)$ leaving node~$i$, we must have $j \in S$. To prove this claim, let us assume that node~$i$ has an edge $(i,j)$ to another node $j \in V - S$. By Lemma~\ref{lem:acyclic}, we know that in this case we must have $\delta_i \geq \delta_j$. However, since we assumed that node~$i$ has the lowest cost in $V-S$, then $\delta_i \leq \delta_j$ and such an edge could only exist if $\delta_i = \delta_j$. By Lemma~\ref{lem:extra}, we know that if $\delta_i = \delta_j$ then the link $(i,j)$ is redundant, and we can safely remove it from the multirate anypath without changing its cost. As a result, for every edge $(i,j)$ we must have $j \in S$. Fig.~\ref{fig:saf_proof} shows this situation where node~$i$ only has links to nodes in~$S$.

Additionally, we claim that the nodes in~$S$ were settled in ascending order of cost. That is, if $\delta_j < \delta_k$ then node~$j$ was settled before node~$k$. Since node~$i$ has the lowest cost to~$d$ among all nodes in $V-S$, settling $s$ before~$i$ implies that~$s$ is settled ``out of order.'' For the purpose of contradiction, let~$s$ be the first node settled out of order. 

We now claim that $D_i = \delta_i$ at the time $s$ is inserted into~$S$. To prove this claim, notice that $K \subseteq S$. Since $s$ is the first node for which $D_s \neq \delta_s$ when it is added to $S$, then we must have $D_k = \delta_k$, for every $k \in K$. Let $\phi_i \subseteq K$ be the forwarding set used in the shortest multirate anypath from~$i$ to~$d$ using the optimal transmission rate $\rho \in R$. 
By Lemma~\ref{lem:order}, $\phi_i$~is composed of the neighbors of~$i$ with the lowest cost to~$d$. Assume that $\phi_i = \{1,2,\ldots,j\}$ with $\delta_1 \leq \delta_2 \leq \ldots \leq \delta_j$. Since $s$ is the first out-of-order node, we know that the nodes in~$S$ were settled in order. Therefore, node~$1$ was settled before node~$2$, which was settled before node~$3$, and so on. At the time node~$1$ is settled, the forwarding set $F_i^{\,(\rho)}$ is initialized to $F_i^{(\rho)}~=~\{1\}$. When node~$2$ is settled, there is no need to check the forwarding set~$\{2\}$. By Lemma~\ref{lem:order}, this forwarding set is never optimal so we just check the set~$\{1,2\}$. By Lemma~\ref{lem:fset}, using $\{1,2\}$ always provides a lower cost than using just~$\{1\}$. The forwarding set is then updated to~$F_i^{\,(\rho)} = \{1,2\}$. The same procedure is repeated for each settled node, until we finally have $F_i^{\,(\rho)} = \phi_i = \{1,2,\ldots,j\}$. At this time, we also have $D_i^{(\rho)} = \delta_i$, which triggers the update $D_i = D_i^{(\rho)} = \delta_i$, $F_i = F_i^{\,(\rho)} = \phi_i$, and $T_i = \rho$. Once $D_i$ is equal to the lowest cost~$\delta_i$, it does not change anymore and we have $D_i = \delta_i$ at the time $s$ is inserted into $S$.

We can now prove the theorem with two contradictions. Since node~$i$ occurs after node~$s$ in the shortest multirate anypath to~$d$, by Lemma~\ref{lem:acyclic} we have $\delta_i \leq \delta_s$. In addition, we must also have $\delta_s \leq D_s$ because $D_s$ is never smaller than~$\delta_s$. Since both $i$ and $s$ are in $V - S$ and node~$s$ was chosen as the one with the minimum cost from~$Q$, then we must have $D_s \leq D_i$ and $\delta_i \leq \delta_s \leq D_s \leq D_i$. From our previous claim, we know that $D_i = \delta_i$ and therefore $D_i = \delta_i \leq \delta_s \leq D_s \leq D_i$, from which we have
$D_i = \delta_i = \delta_s = D_s$. 
As a result, $s$ is not settled out of order since~$i$ has the lowest cost in~$V-S$ and $\delta_s = \delta_i$. From this we conclude that the nodes in~$S$ are settled in ascending order of cost. Additionally, we also have $D_s = \delta_s$ at the time $s$ is added to~$S$, which contradicts our initial choice of~$s$. We conclude therefore that for each node~$s \in V$ we have $D_s = \delta_s$ at the time $s$ is added to~$S$.
\end{IEEEproof}

\begin{thm}
\label{thm:abf_correctness}
{\it
Optimality of the MABF algorithm.

Let $G = (V,E)$ be a weighted directed graph and let $d$ be the destination. After running the Multirate Anypath Bellman-Ford algorithm on $G$, we have $D_i = \delta_i$ for every node $i \in V$.}
\end{thm}

\begin{IEEEproof}
We prove this theorem by induction on $t$, the iteration number. Let $h_i$ be the number of hops of the longest path from~$i$ to~$d$. We show that, after the $t$-th iteration, we have $D_i = \delta_i$ for every node with $h_i \leq t$. Intuitively, the algorithm works from the destination backwards to the source in an expanding-ring fashion, settling at each iteration the nodes one hop further away from the destination. Since we can not have paths with more than $|V|-1$ links, we are guaranteed to have $D_i = \delta_i$ for every node $i \in V$ after $|V|-1$ iterations. The induction proof now follows.

{\bf Basis.} For $t=0$, the only node with $h_i \leq 0$ is the destination~$d$ itself. We have from the initialization that $D_d = \delta_d = 0$.

{\bf Inductive step.} Assuming that after the $t$-th iteration we have $D_i = \delta_i$ for every node with $h_i \leq t$, we want to show that after the $(t+1)$-th iteration we have $D_i = \delta_i$ for every node with $h_i \leq t+1$.
At the $(t+1)$-th iteration, a node~$i$ with $h_i = t+1$ calculates its cost~$D_i^{(r)}$ after checking the forwarding sets $\{1\},\{1,2\},\ldots,\{1,2,\ldots,n\}$ in this order, with $\delta_1 \leq \delta_2 \leq \ldots \leq \delta_n$ being the costs of the neighbors. Since $h_i = t+1$, every neighbor~$j$ in the optimal forwarding set must necessarily have $h_j \leq t$, and we know from the induction hypothesis that $D_j = \delta_j$. From Lemmas~\ref{lem:order} and~\ref{lem:fset}, this strategy is guaranteed to converge to the optimal forwarding set. As a result, we must have $D_i^{(r)} = \delta_i^{(r)}$ for every rate $r \in R$ at the end of the $(t+1)$-th iteration. Therefore, after selecting the best rate we have $D_i = \delta_i$.
\end{IEEEproof}

\section{Experimental Results}
\label{sec:results}

\begin{figure}[ht!]
\centering
\includegraphics[width=.48\textwidth]{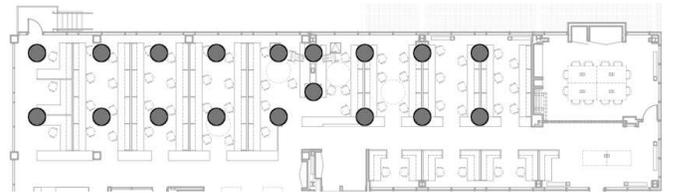}
\caption{The location of nodes in the testbed, in an approximate 2x9 grid.}
\label{fig:testbed-nodes}
\end{figure}

We evaluated the proposed multirate algorithm using an 18-node 802.11b indoor testbed. Each node is a Stargate microserver~\cite{stargate} equipped with an Intel 400-MHz Xscale PXA255 processor, 64 MB of SDRAM, 32 MB of Flash, and an SMC EliteConnect SMC2532W-B PCMCIA 802.11b wireless network card using the Prism2 chipset. This card has a maximum transmission power of 200~mW. 
The nodes of the testbed are distributed over the ceiling of the Center for Embedded Networked Sensing (CENS) at UCLA. The nodes are located in an approximate 2x9 grid and roughly five to ten meters apart from each other. Fig.~\ref{fig:testbed-nodes} depicts the location of the nodes in the testbed. Each node is equipped with a 3-dB omni-directional rubber duck antenna for the wireless communication. In order to emulate a wireless mesh network with multiple hops, we use a 30-dB SA3-XX attenuator between the wireless interface and its antenna. The attenuator weakens the signal during both the transmission and the reception of a frame, emulating a large distance between nodes. For 11~Mbps, we have paths of up to 8 hops between each pair of nodes, with 3.1 hops on average. For 1~Mbps, we have a longer transmission range, which reduces the maximum path length to 3 hops, with an average of 1.5 hops between each pair of nodes.

We use the testbed to measure the delivery ratio of each link at different transmission rates. For that purpose, each node broadcasts one thousand 1500-byte packets and later on we collect the number of received packets at neighbor nodes. We repeat this process for 1, 2, 5.5, and 11~Mbps to have a link estimate for each transmission rate. We use the Click toolkit~\cite{kohler00} and a modified version of the MORE software package~\cite{chachulski07b} for the data collection. Our implementation is capable of sending and receiving raw 802.11 frames by using the wireless network interface in monitor mode. We modified the HostAP Prism driver~\cite{hostap} for Linux in order to allow not only 802.11 frame overhearing but also frame injection while in monitor mode. In addition, we extended the HostAP driver to enable it to control the transmission rate of each 802.11 frame sent. The Click toolkit tags each frame with a selected transmission rate and this information is then passed along to the driver. For each frame, our modification reads the information tagged by Click and notifies the wireless interface firmware about the specified transmission rate. 

\begin{figure}[ht!]
\centering
\includegraphics[width=.4\textwidth]{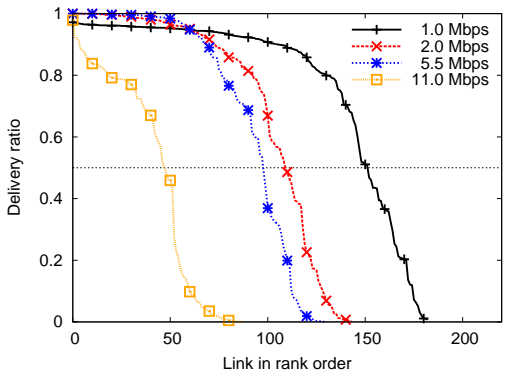}
\caption{The delivery ratio of the testbed links for each transmission rate. The data points for each curve are placed in order from largest to smallest (i.e., in rank order). As the rate increases, fewer links are available and thus path diversity decreases.}
\label{fig:testbed-delivery}
\end{figure}

Fig.~\ref{fig:testbed-delivery} shows the distribution of the delivery ratio of each link in the testbed at different 802.11b transmission rates. Every node pair contributes with two links in the graph, one for each direction. Links of each rate are placed in order from largest to smallest (i.e., in rank order). The points of each curve are sorted separately, and therefore the delivery ratios of a given x-value are not necessarily from the same link. In wireless mesh networks, higher transmission rates usually have shorter transmission ranges and therefore a lower network connectivity. We can see this behavior in Fig.~\ref{fig:testbed-delivery}. As the transmission rate increases, we can see that we have fewer links available and therefore less path diversity between nodes. For instance, as shown by the dashed horizontal line, the number of links with a delivery ratio higher than 50\% is 151 at 1~Mbps, 109 at 2~Mbps, 95 at 5.5~Mbps, and only 47 at 11~Mbps. With fewer paths available at higher rates, we have an interesting tradeoff for multirate anypath routing. With a lower transmission rate, we have more path diversity and a shorter number of hops to traverse, but also a lower throughput. On the other hand, a higher rate results in a higher throughput, but also in less path diversity and a larger number of hops. Our algorithm explores this tradeoff and selects the optimal transmission rate and forwarding set for every node.

\begin{figure}[ht!]
\centering
\includegraphics[width=.4\textwidth]{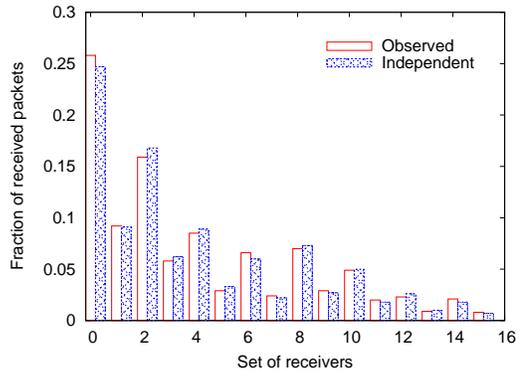}
\caption{(a) Distribution of frame receptions at four neighbors. For four neighbors, we have $2^4 = 16$ subsets and each one represents a different set of neighbors who correctly received the frame. Packet reception at different neighbors is independent for light load regimes.}
\label{fig:independence}
\end{figure}

Fig.~\ref{fig:independence} shows the results of an experiment we conducted to test the independence of receivers. In our experiment, a node individually broadcasts 500,000 data frames at 11~Mbps to four neighbors and each frame has 1500 bytes. The x-axis represents the 16 possible set of receivers for the frame (i.e., set 0 corresponds to the frame being lost by all neighbors and set 15 corresponds to every neighbor correctly receiving the frame). The y-axis represents the fraction of packets received by each set. The ``observed'' histogram is directly derived from the data. The ``independent'' histogram is derived by assuming that the loss probability at each receiver is independent of each other, so it is calculated simply by multiplying the respective probabilities of each individual receiver. We can see that both functions are pretty close, indicating that the delivery ratios of each receiver are loosely correlated in light load regimes. This experiment was repeated for other nodes in the testbed and a similar behavior was observed, which is also consistent with other studies~\cite{miu05,reis06}. We use this observation to derive our next results.

\begin{figure*}[ht!]
\centering
\subfigure[] {
  \includegraphics[width=.4\textwidth]{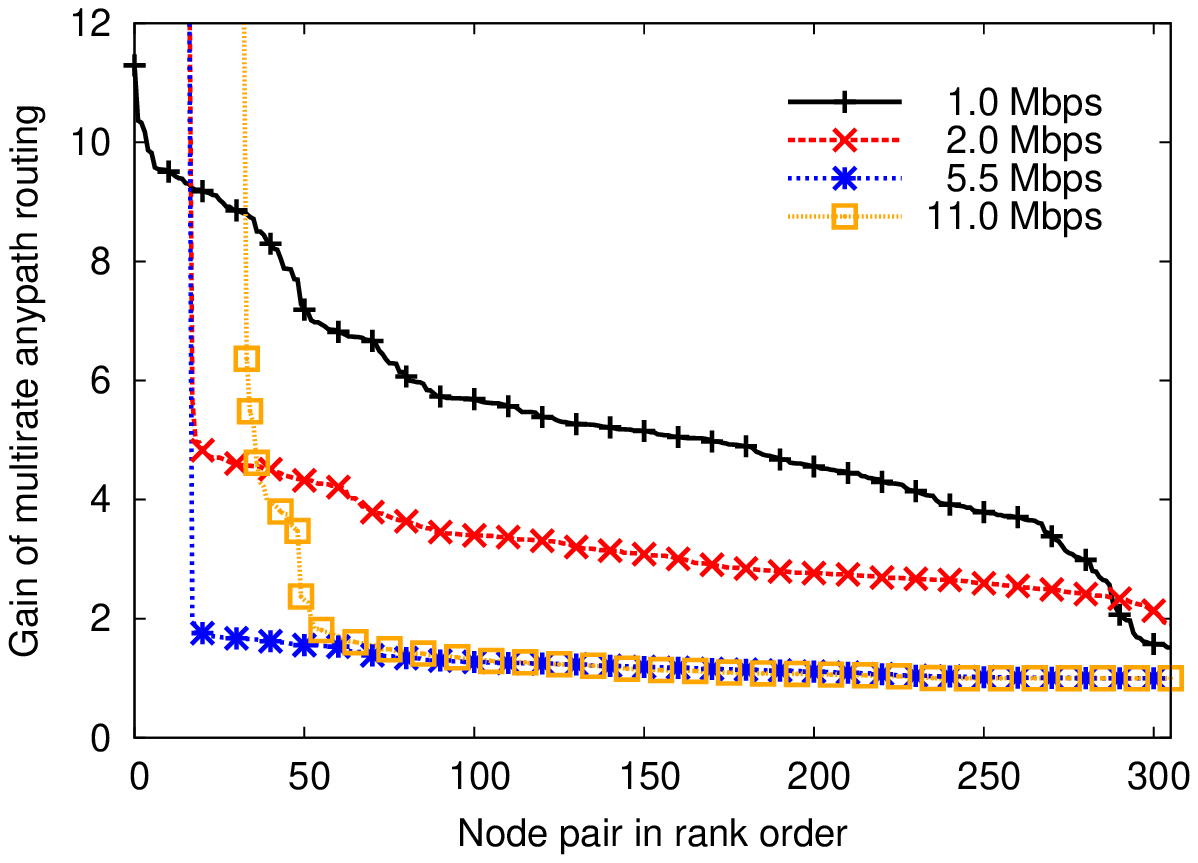}
  \label{fig:gain_ap}
}
\subfigure[] {
  \includegraphics[width=.4\textwidth]{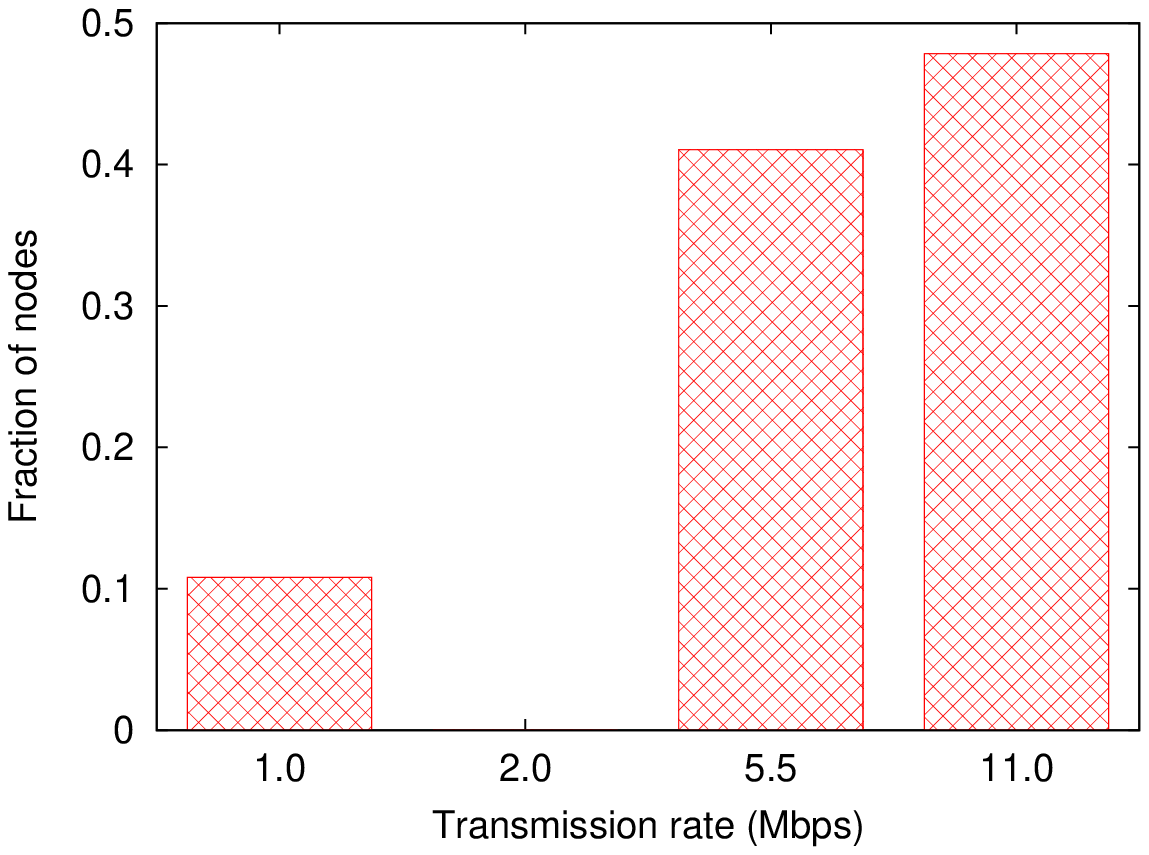}
  \label{fig:rates_pdf}
}
\caption{Results of the SMAF algorithm for the wireless testbed. (a) Gain of multirate over single-rate anypath routing. For each node pair, we indicate in the y-axis how many times multirate anypath routing is better than single-rate anypath. (b) Histogram of the transmission rate chosen by each node. Optimal transmission rates are not concentrated at any particular rate, indicating that a single-rate algorithm can not perform as well as a multirate algorithm.}
\label{fig:results}
\end{figure*}

The shortest multirate anypath {\it always} has an equal or lower cost than the shortest single-rate anypath. Otherwise, we would have a contradiction since we can find another multirate anypath (i.e., the single-rate anypath) with a lower cost to the destination. It is important, however, to quantify how much better multirate anypath routing is over single-rate anypath. For this purpose, we calculate the gain of multirate over single-rate anypath. We define the gain of a given source-destination pair as the ratio between the single-rate anypath cost and the multirate anypath cost between these two nodes. This metric is then a multiplicative factor representing how much longer the end-to-end transmission time is at single-rate anypath routing when compared to multirate anypath routing. 

Fig.~\ref{fig:gain_ap} shows the distribution of this gain for every pair of nodes in the network. Each curve represents the gain over single-rate anypath routing at a fixed rate. We see that the end-to-end transmission time with multirate anypath routing is at least 50\% and up to 11.3 times shorter than with single-rate anypath routing at 1~Mbps, with an average gain of~5.4. For higher rates, we also see an interesting behavior depicted by the vertical lines. These lines indicate that several node pairs have an {\it infinite gain}. The infinite gain occurs because these nodes can not talk to each other at that particular rate due to the poor link quality; the network therefore becomes disconnected. We have 17 (5.6\%) node pairs that can not reach each other at both 2 and 5.5~Mbps, and 33 (10.8\%) node pairs out of reach at 11~Mbps. For the network to be connected, we must then either use a lower rate (i.e., 1~Mbps) for the whole network at the cost of a lower throughput or use multirate anypath routing. For 2~Mbps, if we remove the node pairs with infinite gains, we have a gain of at least 91\% and up to~5.6, with an average of~3.2. For 5.5~Mbps, we have a gain up to 2.0, with an average of 22\%. Finally, for 11~Mbps, we have a gain up to 6.4, with an average of 80\%.

Fig.~\ref{fig:rates_pdf} shows the reason why multirate always performs better than single-rate anypath routing. In this graph we show the distribution of the optimal transmission rates selected by each node to reach every other node. We can see that the optimal transmission rates are not concentrated at a single rate, but rather distributed over several possibilities. We have 10.8\% of node pairs using 1~Mbps, 41\% using 5.5~Mbps, and 47\% using 11~Mbps as the optimal rate. Interestingly enough, no node pair selected 2~Mbps as the optimal rate since it was more beneficial to use another rate instead. If these rates were 100\% concentrated at a particular rate, then multirate and single-rate anypath routing would have the same cost. This assumption, however, does not hold in practice, and therefore multirate anypath routing always has a higher performance, sometimes manyfold higher as shown in Fig.~\ref{fig:gain_ap}, than single-rate anypath routing.

\section{Related Work}
\label{sec:related}

Most of the work in anypath routing focuses on using a single transmission rate. The following works are all single-rate anypath routing schemes. 

Zorzi and Rao~\cite{zorzi03} use a combination of opportunistic and geographic routing in a wireless sensor network. The authors assume that sensor nodes are aware of their locations and this information is used for routing. The forwarding set of a given node is composed of the neighbors which are physically closer to the destination. Packets are broadcast and neighbors in the set forward the packet respecting the relay priority explained in Section~\ref{sec:anypath}. As an advantage, this routing procedure does not need any sort of route dissemination over the network. Using just the physical distance as the routing metric, however, may not be the best approach since it does not take link quality into account. We introduce the EATT routing metric that takes not only the link quality but also the multiple transmission rates into account during route calculation.

Ye {\it et al.}~\cite{ye05} present another single-rate opportunistic routing protocol for sensor networks. The key idea is that each packet carries a credit which is initially set by the source and is reduced as the packet traverses the network. Each node also maintains a cost for forwarding a packet from itself to the destination, and nodes closer to the destination have smaller costs. Packets are sent in broadcast and a neighbor node forwards a received packet only if the credit in the packet is high enough. Just before forwarding the packet, its credit is reduced according to the node cost; therefore, more credits are consumed as the packet moves away from the shortest path. A mesh around the shortest path is then created on the fly for each packet. Yuan~{\it et~al.}~\cite{yuan05} use a similar idea for wireless mesh networks. Although packet delivery is improved, this routing scheme increases overhead since it is based on a controlled flooding mechanism. Therefore, robustness comes at the cost of duplicate packets. In our proposal, a packet is forwarded by a single neighbor in the forwarding set and a MAC mechanism, such as the one proposed by Jain and Das~\cite{jain08}, is in place to guarantee that no duplicate packets occur in the network.

Biswas and Morris~\cite{biswas05a} designed and implemented ExOR, an opportunistic routing protocol for wireless mesh networks. ExOR follows the same guidelines of single-rate anypath routing explained in Section~\ref{sec:anypath}. Basically, a node forwards a batch of packets, and each neighbor in the forwarding set waits its turn to transmit the received packets. The authors implement a MAC scheduling scheme to enforce the relay priority in the forwarding set. As a result, a node only forwards a packet if all higher priority nodes failed to do so. The authors show that opportunistic routing increases throughput by a factor of two to four compared to single-path routing. Our results go beyond that and show that an even better performance can be achieved with multirate anypath routing. Additionally, in our design, each packet is routed independently without storing any per-batch state at intermediate routers.

Chachulski {\it et al.}~\cite{chachulski07b} introduce MORE, a routing protocol which uses both opportunistic routing and network coding to increase the network end-to-end throughput. Upon the receipt of a new packet, a node encodes it with previously received packets and then broadcasts the coded packet. 
Results show that MORE allows a higher throughput than ExOR and single-path routing. Network coding, however, requires routers to store previous packets in order to code them with future packets, adding significant storage and processing overhead to the forwarding process. Furthermore, the authors only focus on opportunistic routing with a single transmission rate. 
Our results indicate that performance could be further improved with multirate anypath routing. An analysis of multirate anypath routing and network coding is also an open problem and an interesting topic for future work. 

Besides using a single bit rate, the above-mentioned systems also do not have a systematic approach for selecting the forwarding set for a given destination. The selection is commonly based on the heuristic that if a neighbor has a smaller ETX cost to the destination, then it should be in the forwarding set. However, the ETX is a single-path metric and does not correctly represent the true cost when using anypath routing. To our knowledge, Zhong~{\it et~al.}~\cite{zhong06a} was the first to propose the expected anypath number of transmissions (EATX) metric described in Section~\ref{sec:anypath}, which was also used in~\cite{chachulski07b,dubois-ferriere10}. The authors propose an algorithm for forwarding set selection in~\cite{zhong07}, but this algorithm is not proven optimal.

Dubois-Ferrière {\it et al.}~\cite{dubois-ferriere10} introduced a shortest anypath algorithm capable of finding optimal forwarding sets. The authors generalize the well-known Bellman-Ford algorithm for anypath routing and prove its optimality. Performance tests in a wireless sensor network show that anypath routing significantly reduces the required number of transmissions from a node to the sink. The running time of this algorithm, however, is exponential. Chachulski~\cite{chachulski07b} presents a generalization of Dijkstra's algorithm for anypath routing that is similar to the one we independently derived in Section~\ref{sec:single-rate}, but the author does not provide any proof of optimality. Both of these algorithms, however, are designed for networks using a single transmission rate. Instead, our algorithms in Section~\ref{sec:multi-rate} generalize anypath routing for multiple rates, giving nodes the ability to choose both the best rate and the best forwarding set to a particular destination. We also provide the proof of optimality for our algorithms. As a result, the optimality of the single-rate Dijkstra's algorithm in~\cite{chachulski07b} is also proved since this is a special case of our algorithm. In a concurrent work, Lu and Wu~\cite{lu09} propose a routing algebra for opportunistic routing and also introduce similar algorithms for single-rate anypath routing.

More recently, multiple transmission rates have been addressed in opportunistic routing. Radunovic {\it et al.}~\cite{radunovic08} presents an optimization framework to derive routing, scheduling, and rate adaptation schemes. Zeng~{\it et~al.}~\cite{zeng08} presents a linear-programming formulation to optimize the end-to-end throughput of opportunistic routing, considering multiple rates and transmission conflict graphs. However, in both cases the authors try to optimize several components simultaneously, and therefore the posed problem becomes NP-hard. Heuristics are then applied to find a solution, which is not necessarily optimal and may not be easily implemented. Instead, we focus on the shortest multirate anypath problem and provide an optimal solution for it in polynomial time.

\section{Conclusions}
\label{sec:conclusions}

In this paper we introduced multirate anypath routing, a new routing paradigm for wireless multihop networks. We provided a solution to integrate opportunistic routing and multiple transmission rates. The available rate diversity imposes several new challenges to routing, since transmission range and delivery ratios change with the transmission rate. Given a network topology and a destination, we set out to find both a forwarding set and a transmission rate for every node, such that their cost to the destination is minimized. We pose this as the {\it shortest multirate anypath problem}. Finding the rate and forwarding set that jointly optimize the cost from a node to a given destination was previously considered an open problem. To solve it, we introduced the EATT routing metric as well as the Shortest Multirate Anypath First (SMAF) and the Multirate Anypath Bellman-Ford (MABF) algorithms and presented a proof of their optimality. Our algorithms have roughly the same complexity as regular shortest-path algorithms, being easy to implement in current routing protocols.

We conducted experiments in a 18-node 802.11b testbed to evaluate the performance of multirate over single-rate anypath routing. Our main findings are: (1) when the network uses a single bit rate, it may become disconnected since some links may not work at the selected rate; (2) multirate outperforms single-rate 11-Mbps anypath routing by 80\% on average and up to a factor of 6.4 while still maintaining full connectivity; (3) multirate also outperforms single-rate 1-Mbps anypath routing by a factor of 5.4 on average and up to a factor of~11.3; (4) the distribution of the optimal transmission rates is not concentrated at any particular rate, corroborating the assumption that nodes in single-rate anypath routing usually do not transmit at their optimal rates. 

\section*{Acknowledgments}
This work was done in part while the first author was visiting the Ecole Polytechnique Fédérale de Lausanne (EPFL) in the summer of 2007. We would like to thank Martin Vetterli for hosting the first author at EPFL and introducing anypath routing to him. We thank Deborah Estrin for her help and discussions over the years and for use of the CENS testbed. We are grateful to Martin Lukac for his help with the testbed. Any opinions, findings, and conclusions or recommendations expressed in this material are those of the authors and do not necessarily reflect the views of the National Science Foundation and of Alcatel-Lucent. 

\bibliographystyle{IEEEtran}
\bibliography{report}

\appendix
\section{Proofs of the Lemmas}
\label{app:proofs}

{\it Lemma~\ref{lem:comparison}: For a fixed transmission rate, let $D_i$ be the cost of a node~$i$ via forwarding set~$J$ and let $D_i'$ be the cost via forwarding set $J' = J \cup \{k\}$, where $D_k \geq D_j$ for every node $j \in J$. We have $D_i' \leq D_i$ if and only if $D_i \geq D_k$.}
\begin{IEEEproof}
Assume that $D_1 \leq D_2 \leq \ldots \leq D_{k-1} \leq D_k$ and $J=\{1,2,\ldots,k-1\}$. Let $D_i = d_{iJ} + D_J$ be the cost from node~$i$ using the forwarding set~$J$. 
From~(\ref{eq:D_J_def}), the remaining anypath cost is generally defined as $D_J = \sum_{j \in J} w_{ij}D_j$, with $\sum_{j \in J} w_{ij} = 1$,
where the weight $w_{ij}$ is the probability of using node~$j$ as the relay, given that at least one node in~$J$ received the packet.
Let $D_i' = d_{iJ'} + D_{J'}$ be this cost via $J'=J \cup \{k\}$. 
First, we write $D_{J'}$ in terms of $D_J$ as
\begin{equation}
\label{eq:aggregated} D_{J'} = \frac{p_{iJ}}{p_{iJ'}}D_J + \left(1-\frac{p_{iJ}}{p_{iJ'}}\right)D_k,
\end{equation}
where $p_{iJ}/p_{iJ'}$ 
scales the weights in $D_J$ to account for the new forwarding set~$J'$ by changing the probability of the condition in the denominator of~$w_{ij}$. The probability in the numerator of~$w_{ij}$ does not change. The $(1-p_{iJ}/p_{iJ'})$ factor is the conditional probability that node~$k$ is the relay, given that at least one node in $J'$ received the packet.
An interesting result from~(\ref{eq:aggregated}) is that 
we can see the forwarding set $J$ as an ``aggregated node'' with delivery ratio~$p_{iJ}$ and cost~$D_J$.

We now show that if $D_i \geq D_k$, then $D_i' \leq D_i$ as follows
\setlength{\arraycolsep}{0.0em}%
\begin{eqnarray}
\nonumber D_i &{}\geq{}& D_k \\
\nonumber \left(1-\frac{p_J}{p_{J'}}\right)\left(d_{iJ} + D_J\right) &{}\geq{}& \left(1-\frac{p_J}{p_{J'}}\right)D_k \\
\nonumber d_{iJ} + D_J &{}\geq{}& \frac{p_J}{p_{J'}}d_{iJ} + \frac{p_J}{p_{J'}}D_J + \left(1-\frac{p_J}{p_{J'}}\right)D_k\\
\nonumber D_i &{}\geq{}& d_{iJ'} + D_{J'} \\
\label{eq:D_i_and_D_i'} D_i &{}\geq{}& D_i'.
\end{eqnarray}
\setlength{\arraycolsep}{5pt}%
To show that if $D_i' \leq D_i$ then $D_i \geq D_k$, we just take~(\ref{eq:D_i_and_D_i'}) in the reverse order. Consequently, if $D_i > D_k$, it is better to use the forwarding set $J'= J \cup \{k\}$ instead of~$J$, since the cost~$D_i'$ via~$J'$ is always lower than~$D_i$ via~$J$.
\end{IEEEproof}

\medskip
{\it Lemma~\ref{lem:acyclic}: The lowest cost~$\delta_i$ of a node~$i$ is always larger than or equal to the lowest cost $\delta_j$ of any node~$j$ in the optimal forwarding set $\phi_i$. That is, we have $\delta_i \geq \delta_j$ for all $j \in \phi_i$.}
\begin{IEEEproof}
Let $\phi_i = \{1,2,\ldots,k\}$ be the optimal forwarding set with $\delta_1 \leq \delta_2 \leq \ldots \leq \delta_k$ and let $D_i \geq \delta_i$ be the cost via the suboptimal forwarding set $J = \{1,2,\ldots,k-1\}$ with the same transmission rate. From Lemma~\ref{lem:comparison}, we know that if $\delta_i \leq D_i$, then $D_i \geq \delta_k$. From this, we show that $\delta_i \geq \delta_k$ as follows (assume $J'$ is the optimal forwarding set $\phi_i$)
\setlength{\arraycolsep}{0.0em}%
\begin{eqnarray}
\nonumber D_i &{}\geq{}& \delta_k \\
\nonumber \left(\frac{p_J}{p_{J'}}\right)\left(d_{iJ} + D_J\right) &{}\geq{}& \left(\frac{p_J}{p_{J'}}\right)\delta_k \\
\nonumber \frac{p_J}{p_{J'}}d_{iJ} + \frac{p_J}{p_{J'}}D_J &{}\geq{}& \delta_k - \left(1-\frac{p_J}{p_{J'}}\right)\delta_k \\
\nonumber d_{iJ'} + \frac{p_J}{p_{J'}}D_J + \left(1-\frac{p_J}{p_{J'}}\right)\delta_k &{}\geq{}& \delta_k \\
\nonumber d_{iJ'} + D_{J'} &{}\geq{}& \delta_k \\
\delta_i &{}\geq{}& \delta_k.
\end{eqnarray}
\setlength{\arraycolsep}{5pt}%
Since $\delta_k$ is the highest cost in the optimal forwarding set~$\phi_i$,  then we know that if $\delta_i \geq \delta_k$, then $\delta_i \geq \delta_j$ for all $j \in \phi_i$.
\end{IEEEproof}

\medskip
{\it Lemma~\ref{lem:extra}: For any transmission rate, if a node~$i$ uses a node~$k$ in its optimal forwarding set~$\phi_i$ and $\delta_i = \delta_k$, we can safely remove~$k$ from~$\phi_i$ without changing~$\delta_i$. The link $(i,k)$ is said to be ``redundant.''}
\begin{IEEEproof}
By Lemma~\ref{lem:acyclic}, we have $\delta_i \geq \delta_j$, for all $j\in\phi_i$. Since $\delta_i = \delta_k$, we also know that $\delta_k$ is the highest cost in the forwarding set. Let $J' = \phi_i = \{1,2,\ldots,k\}$ be the optimal forwarding set with $\delta_1 \leq \delta_2 \leq \ldots \leq \delta_k$ and let $D_i = d_{iJ} + D_J$ be the cost from node~$i$ via $J = \{1,2,\ldots,k-1\}$. We now show that if $\delta_i = \delta_k$, then $D_i = \delta_i$ as follows
\setlength{\arraycolsep}{0.0em}%
\begin{eqnarray}
\nonumber \delta_i &{}={}& \delta_k \\
\nonumber d_{iJ'} + D_{J'} &{}={}& \delta_k \\
\nonumber d_{iJ'} + \frac{p_{iJ}}{p_{iJ'}}D_J + \left(1-\frac{p_{iJ}}{p_{iJ'}}\right)\delta_k &{}={}& \delta_k \\
\nonumber d_{iJ'} + \frac{p_{iJ}}{p_{iJ'}}D_J &{}={}& \delta_k - \left(1-\frac{p_{iJ}}{p_{iJ'}}\right)\delta_k \\
\nonumber d_{iJ'} + \frac{p_{iJ}}{p_{iJ'}}D_J &{}={}& \frac{p_{iJ}}{p_{iJ'}}\delta_k\\
\nonumber \frac{p_{iJ'}}{p_{iJ}}d_{iJ'} + D_J &{}={}& \delta_k\\
\nonumber d_{iJ} + D_J &{}={}& \delta_k\\
D_i &{}={}& \delta_k.
\end{eqnarray}
\setlength{\arraycolsep}{5pt}%
Since $D_i = \delta_k$, the forwarding set~$J$ is also optimal and yields the same cost as $\phi_i$. We say the link $(i,k)$ is ``redundant'' since it does not help to reduce the cost any further.
\end{IEEEproof}

\medskip
{\it Lemma~\ref{lem:order}: If the lowest costs from the neighbors of a node~$i$ to a given destination are $\delta_1 \leq \delta_2 \leq \ldots \leq \delta_n$, then the optimal forwarding set~$\phi_i^{(r)}$ is always of the form $\phi_i^{(r)} = \{1,2,\ldots,k\}$, for some $k \in \{1,2,\ldots,n\}$.}
\begin{IEEEproof}
Define a sequential forwarding set as a set in which neighbors are grouped without any gaps in the cost sequence (i.e., a set of the form $\{1,2,\ldots,l-1,l\}$). Now assume that the optimal forwarding set $J = \phi_i^{(r)}$ is not sequential. Let $\delta_i$ be the cost of node $i$ via $J$, and let $l$ be the neighbor with the highest cost in $J$. Since $J$ is not sequential, there must be at least one node $k \not \in J$ such that $\delta_k \leq \delta_l$.
Without loss of generality, let $k < l$ be the lowest cost neighbor that is not in~$J$, and let $D_i'$ be the cost of~$i$ via $J' = J \cup \{k\}$. If $J$ is the optimal forwarding set, then $\delta_i \leq D_i'$ and 
\begin{equation}
\label{eq:optset} \left(1-\frac{p_{iJ}}{p_{iJ'}}\right)\delta_i \leq D_i' - \frac{p_{iJ}}{p_{iJ'}}\delta_i = D_{J'} - \frac{p_{iJ}}{p_{iJ'}}D_J.
\end{equation}
where the last equality holds because $d_{iJ'} = d_{iJ}p_{iJ}/p_{iJ'}$. By Lemma~\ref{lem:acyclic}, we know that $\delta_i \geq \delta_l$ and from~(\ref{eq:optset}) we have
\setlength{\arraycolsep}{0.0em}%
\begin{eqnarray}
\nonumber \left(1 - \frac{p_{iJ}}{p_{iJ'}}\right)\delta_l &{}\leq{}& D_{J'} - \frac{p_{iJ}}{p_{iJ'}}D_J\\
\nonumber  &{}={}& \sum_{j \in J'} w_{ij}'\delta_j - \frac{p_{iJ}}{p_{iJ'}} \sum_{j \in J} w_{ij}\delta_j\\
\nonumber  &{}={}& w_{ik}'\delta_k - \sum_{j \in J} \left(\frac{p_{iJ}}{p_{iJ'}} w_{ij} - w_{ij}'\right)\delta_j\\
\nonumber  &{}\leq{}& w_{ik}'\delta_k - \sum_{j \in J} \left(\frac{p_{iJ}}{p_{iJ'}} w_{ij} - w_{ij}'\right)\delta_k\\
\label{eq:optset2} &{}={}& \left(1 - \frac{p_{iJ}}{p_{iJ'}}\right)\delta_k,
\end{eqnarray}
\setlength{\arraycolsep}{5pt}%
where we replace each $\delta_j$ with $\delta_k$ since $w_{ij}p_{iJ}/p_{iJ'} - w_{ij}' = 0$ for $\delta_j < \delta_k$ and non-negative for $\delta_j \geq \delta_k$. As a result, $\delta_k \geq \delta_l$. However, since by definition $\delta_k \leq \delta_{k+1} \leq \ldots \leq \delta_{l-1} \leq \delta_l$, then we must have $\delta_k = \delta_{k+1} = \ldots = \delta_{l-1} = \delta_l$, which contradicts our initial assumption that the optimal forwarding set is not sequential.
\end{IEEEproof}

\bigskip
\medskip
{\it Lemma~\ref{lem:fset}: For a given transmission rate $r \in R$, assume that $\phi_i^{(r)} = \{1,2,\ldots,k\}$ with costs $\delta_1 \leq \delta_2 \leq \ldots \leq \delta_k$. If $D_i^j$ is the cost from node~$i$ using transmission rate~$r$ via forwarding set $\{1,2,\ldots,j\}$, for $1 \leq j \leq k$, then we always have $D_i^1 \geq D_i^2 \geq \ldots \geq D_i^k = \delta_i^{(r)}$.}

\begin{IEEEproof}
We want to prove that the set $J = \{1\}$ yields a higher cost than $J' = \{1,2\}$, which yields a higher cost than $J'' = \{1,2,3\}$ and so on until we get to the optimal forwarding set $\phi_i^{(r)} = \{1,2,\ldots,k\}$. Thus far, we have
\begin{equation}
\label{eq:so_far} \delta_{k-1} \stackrel{(a)}{\leq} \delta_k \stackrel{(b)}{\leq} \delta_i \stackrel{\raisebox{.125cm}{\scriptsize $(c)$}}{=} D_i^k \stackrel{(d)}{\leq} D_i^{k-1}, 
\end{equation}
where $(a)$ and $(c)$ hold by definition, $(b)$ holds by Lemma~\ref{lem:acyclic}, and $(d)$ holds because $\phi_i^{(r)} = \{1,2,\ldots,k\}$ yields the lowest cost to the destination at rate~$r$. 

We now extend this result further for other forwarding sets. We first claim that
\begin{equation}
\delta_{k-2} \stackrel{(a)}{\leq} \delta_{k-1} \leq \delta_k \leq \delta_i = D_i^k \leq D_i^{k-1} \stackrel{(b)}{\leq} D_i^{k-2},
\end{equation}
where $(a)$ holds by definition and $(b)$ holds because of the following argument. By definition, we have $\delta_i = D_i^k \leq D_i^{k-2}$, since $D_i^k$ is the lowest cost to the destination. From~(\ref{eq:so_far}), we then have that $\delta_{k-1} \leq D_i^{k-2}$. Finally, if $\delta_{k-1} \leq D_i^{k-2}$, then $D_i^{k-1} \leq D_i^{k-2}$ by Lemma~\ref{lem:comparison}. 

The same argument can be made recursively until we get
\begin{equation}
\hspace{-0mm} \delta_{1} \leq \ldots \leq \delta_{k-1} \leq \delta_k \leq \delta_i = D_i^k \leq D_i^{k-1} \leq \ldots \leq D_i^{1}, \hspace{-4mm}
\end{equation}
from which we know $D_i^1 \geq D_i^2 \geq \ldots \geq D_i^k$ must be true.
\end{IEEEproof}

\end{document}